\documentclass[useAMS,usenatbib]{mn2e}
\usepackage{graphicx}
\usepackage{latexsym}
\usepackage{psfig}
\usepackage{amssymb,amsmath}
\usepackage{subfig}
\usepackage{caption}

\title{Ionizing feedback from massive stars in massive clusters III: Disruption of partially unbound clouds}

\author[J. E. Dale, B. Ercolano, I.A. Bonnell]{J. E. Dale$^{1}$\thanks{E-mail: dale@usm.lmu.de (JED)}, B. Ercolano$^{1}$, I. A. Bonnell$^{2}$\\
$^{1}$Excellence Cluster `Universe', Boltzmannstr. 2, 85748 Garching, Germany.\\
$^{2}$Department of Physics and Astronomy, University of St Andrews, North Haugh, St Andrews, Fife KY16 9SS}

\begin{document}

\pagerange{\pageref{firstpage}--\pageref{lastpage}} \pubyear{2006}

\maketitle

\label{firstpage}

\def\mnras{MNRAS}
\def\apj{ApJ}
\def\aj{AJ}
\def\aap{A\&A}
\def\apjl{ApJL}
\def\apjs{ApJS}
\def\araa{ARA\&A}
\def\pasp{PASP}

\begin{abstract}
We extend our previous SPH parameter study of the effects of photoionization from O--stars on star--forming clouds to include initially unbound clouds. We generate a set of model clouds in the mass range $10^{4}$-10$^{6}$M$_{\odot}$ with initial virial ratios $E_{\rm kin}/E_{\rm pot}$=2.3, allow them to form stars, and study the impact of the photoionizing radiation produced by the massive stars. We find that, on the 3Myr timescale before supernovae are expected to begin detonating, the fractions of mass expelled by ionizing feedback is a very strong function of the cloud escape velocities. High--mass clouds are largely unaffected dynamically, while lower--mass clouds have large fractions of their gas reserves expelled on this timescale. However, the fractions of stellar mass unbound are modest and significant portions of the unbound stars are so only because the clouds themselves are initially partially unbound. We find that ionization is much more able to create well--cleared bubbles in the unbound clouds, owing to their intrinsic expansion, but that the presence of such bubbles does not necessarily  indicate that a given cloud has been strongly influenced by feedback. We also find, in common with the bound clouds from our earlier work, that many  of the systems simulated here are highly porous to photons and supernova ejecta, and that most of them will likely survive their first supernova explosions.
\end{abstract}

 
\section{Introduction}
The vast majority of stars are formed by the gravitational collapse of regions of giant molecular clouds (GMCs) \citep[e.g.][]{2003ARA&A..41...57L}. Since this process is gravitationally--driven, the regions where star formation occurs must be gravitationally bound and it was long thought that all GMCs themselves must therefore be globally bound \citep[e.g.][]{1979IAUS...84...35S}. However, GMCs convert only a small fraction of their gas reserves to stars and most stars do not appear to live in gravitationally bound systems \citep[e.g.][]{2003ARA&A..41...57L}. It seems obvious, then, that some process must begin shortly after star formation starts in GMCs which rapidly stops the cloud forming stars and gravitationally disrupts both the remaining gas and (usually) the embedded star cluster(s) inside.\\
\indent Stellar feedback in the form of ionizing radiation, winds, jets and supernova explosions are obvious candidate mechanisms for expelling gas from GMCs and have been modelled analytically or numerically by many authors. We will not give a detailed discussion of these mechanisms here but merely refer to the introduction to our preceding paper (\cite{2012MNRAS.424..377D}, hereafter Paper I). Recently, \cite{2012MNRAS.419..841K} showed that, even in the absence of any kind of feedback, the volumes of embedded clusters can be cleared of gas by a combination of accretion and the resulting shrinkage of the clusters. It is hard to see, however, how this process can be responsible for terminating star formation on the scale of a GMC, or unbinding such an object, since it does not actually disperse or destroy the cold natal gas.\\
\indent Whilst several studies have shown that efficient gas removal can disrupt embedded clusters, it is by no means clear that any or all of the stellar feedback mechanisms mentioned above can actually achieve this feat on short enough timescales. Simulations of gas expulsion by O--star photoionization or winds \citep[e.g][]{2005MNRAS.358..291D,2006ApJ...653..361K,2008MNRAS.391....2D,2010ApJ...715.1302V,2011MNRAS.414..321D,2012MNRAS.424..377D} have found that these are rather slow processes, particularly in massive clouds with high escape velocities. Other authors have shown that young clusters may be effectively disrupted by tidal shocks from encounters with passing molecular clouds but this also occurs on a long timescale and leaves the problem of gas removal unsolved \citep[e.g.][]{2006MNRAS.371..793G,2012MNRAS.421.1927K}. \\
\indent The problem of dispersing embedded clusters into the field could be circumvented if the clouds from which they form are never globally bound in the first place. That this may be the case for many or most clouds has been suggested by many authors \citep{2001MNRAS.327..663P,2005ApJ...618..344V,2005MNRAS.359..809C,2008MNRAS.386....3C,2009ApJ...699.1092H,2011MNRAS.413.2935D}. It is quite possible to conceive of a turbulent cloud whose global virial ratio is well in excess of unity, so that it is formally \emph{globally} unbound, but where there are regions of gas formed by converging flows that are \emph{locally} bound and where star formation may proceed \citep{2005MNRAS.359..809C}. This would lead to relatively small numbers of unbound or lightly--bound stars and would naturally result in a low star--formation efficiency \citep{2004MNRAS.347L..36C}, since large fractions of the gas are never likely to become gravitationally unstable. \cite{2008MNRAS.386....3C} performed a series of simulations with a range of virial ratios from $E_{\rm kin}/|E_{\rm grav}|$=0.1 to 10 and found that scaling this ratio allowed arbitrary SFEs to be produced in clouds evolved for one or two dynamical times. This occurred partly because the onset of star formation was delayed in the more unbound clouds, but also because it proceeded at a slower rate. This in turn eases the problem of gas expulsion by removing the need for it to be done quickly, and ensuring that much of the gas will probably disperse of its own accord anyway. In our previous study involving bound clouds, we found that gas--removal by photoionization was a slow and inefficient process in many clouds and its ability to halt star formation and to unbind the stellar content of the clouds was modest. Here, we will see if starting from clouds which are already partially unbound aids photoionization in either of these directions.\\
\indent We describe our numerical techniques in Section 2, discuss our chosen parameter space in Section 3, present our results in Section 4 and our discussion and conclusions follow in Sections 5 and 6 respectively.\\
\section{Numerical methods}
The numerical methods used in this work are identical to those used in Paper I and will only be very briefly described here. We use a well--known variant of the \citep{1990nmns.work..269B} Smoothed Particle Hydrodynamics \citep{1992ARA&A..30..543M} code, which is ideal for studying the evolution of molecular clouds and embedded clusters. In all our simulations, we begin with 10$^{6}$ gas particles.  We use the standard artificial viscosity prescription, with $\alpha=1$, $\beta=2$. Particles are evolved on individual timesteps. The code is a hybrid N--body SPH code in which stars are represented by point--mass sink particles \citep{1995MNRAS.277..362B}. Self--gravitational forces between gas particles are calculated using a binary tree, whereas gravitational forces involving sink--particles are computed by direct summation. Sink particles are formed dynamically and may accrete gas particles and grow in mass. In our simulations of 10$^{5}$ and 10$^{6}$M$_{\odot}$ clouds, the sink particles represent stellar clusters, since the mass resolution is not sufficient to capture individual stars. The accretion radii of the sinks representing clusters are chosen to be always $\lesssim$1$\%$ of the radius of the simulated clouds. Clusters approaching each other to within their accretion radii are merged if they are mutually gravitationally bound. In our 10$^{4}$M$_{\odot}$ and 3$\times10^{4}$M$_{\odot}$ simulations, sink particles represent individual stars. Their accretion radii are set to 0.005pc ($\sim10^{3}$ AU) and mergers are not permitted. In all simulations gravitational interactions of sink particles with other sink particles are smoothed within their accretion radii.\\
\indent We treat the thermodynamics of the neutral gas using a piecewise barotropic equation of state from \cite{2005MNRAS.359..211L}, defined so that $P = k \rho^{\gamma}$, where
\begin{eqnarray}
\begin{array}{rlrl}
\gamma  &=  0.75  ; & \hfill &\rho \le \rho_1 \\
\gamma  &=  1.0  ; & \rho_1 \le & \rho  \le \rho_2 \\
\gamma  &=  1.4  ; & \hfill \rho_2 \le &\rho \le \rho_3 \\
\gamma  &=  1.0  ; & \hfill &\rho \ge \rho_3, \\
\end{array}
\label{eqn:eos}
\end{eqnarray}
and $\rho_1= 5.5 \times 10^{-19} {\rm g\ cm}^{-3} , \rho_2=5.5 \times10^{-15} {\rm g cm}^{-3} , \rho_3=2 \times 10^{-13} {\rm g\ cm}^{-3}$. We discuss our choice and justification of this equation of state in some detail in Paper I, and we also showed that exchanging it for a purely isothermal equation of state had little impact on our results.\\
\begin{figure}
\includegraphics[width=0.45\textwidth]{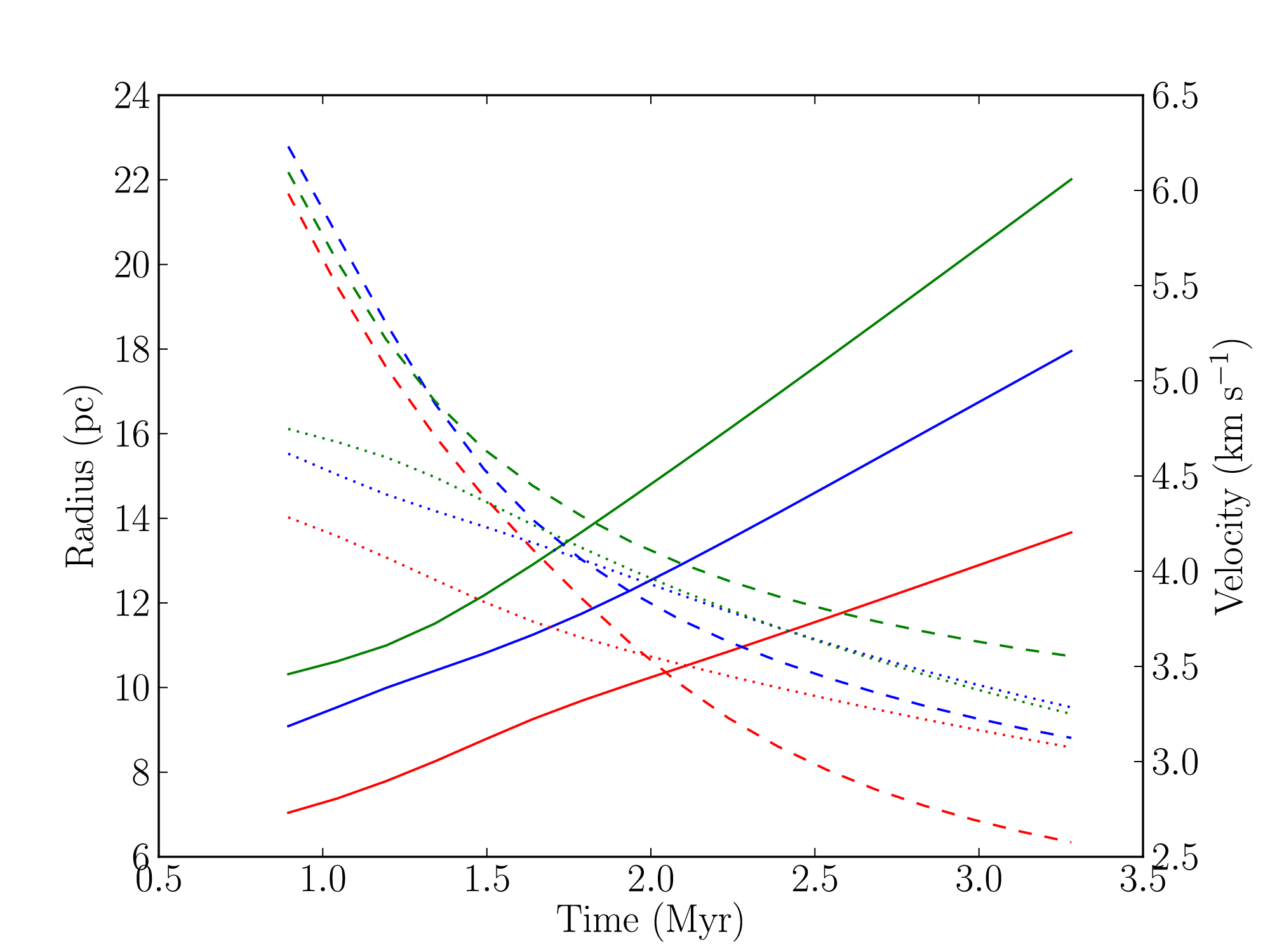}
\caption{Radii (solid lines), velocity dispersions (dashed lines) and escape velocities (dotted lines) plotted at the 50$\%$ (red), 75$\%$ (blue) and 90$\%$ (green) Lagrange radii in Run UF.}
\label{fig:uf_rv}
\end{figure}
\begin{figure}
\includegraphics[width=0.45\textwidth]{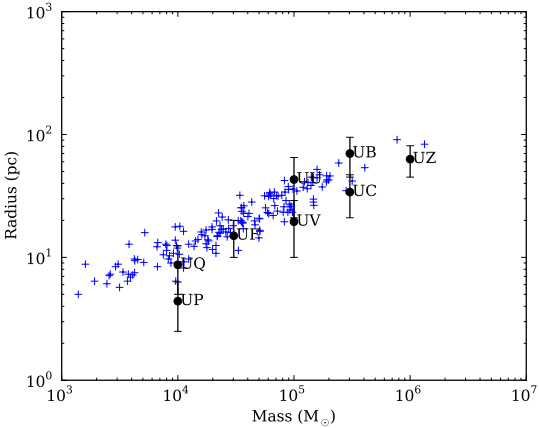}
\caption{Radius--mass parameter space for clouds studied here. Vertical bars represent the range from the initial cloud radius (the lower limits) to the radius at which ionization becomes active in each cloud (upper limits), with black dots being the algebraic mean radius, and `radius' in all cases connotes the 90$\%$ Lagrange radius.}
\label{fig:mr}
\end{figure}
\begin{figure}
\includegraphics[width=0.45\textwidth]{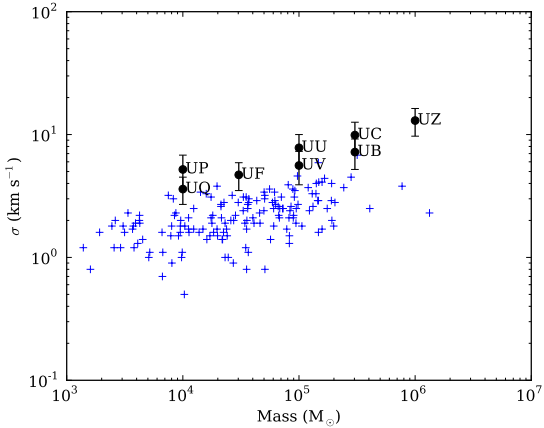}
\caption{Velocity--dispersion--mass parameter space for clouds studied here. Vertical bars represent the range from the initial turbulent velocity dispersion (the upper limits) to the velocity dispersion extant at the time when ionization becomes active in each cloud (lower limits), with black dots being the algebraic mean of these values.}
\label{fig:msigma}
\end{figure}
\begin{figure}
\includegraphics[width=0.45\textwidth]{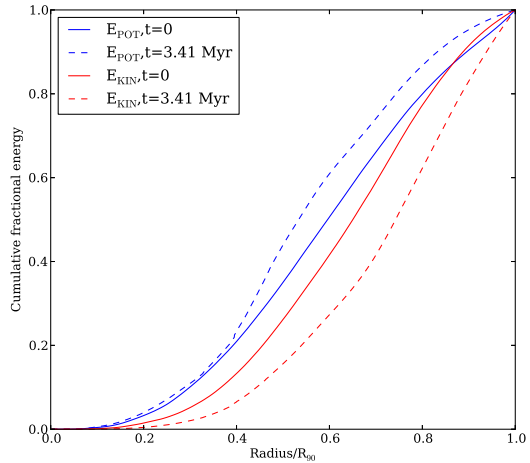}
\caption{Cumulative distribution of kinetic (red) and potential (blue) energies with radius (as a fraction of the 90$\%$ Lagrangian radius) within the Run UF cloud in its initial state (solid lines) and at the epoch when ionization is enabled (dashed lines).}
\label{fig:energy_dist}
\end{figure}
\indent We use the multiple--source photoionization code described in Paper I. The code uses a simple ray--tracing algorithm and a Str\"omgren volume technique to compute the flux of ionizing photons arriving at a given SPH particle and update its ionization state accordingly. The on--the--spot approximation is used and the modified recombination coefficient $\alpha_{\rm B}$ is taken to be $3.0\times10^{-13}$ cm$^{3}$ s$^{-1}$. Fully ionized particles are given a temperature of 10$^{4}$ K, whereas partially ionized particles are given temperatures computed from multiplying their ionization fraction by 10$^{4}$ K. The action of multiple ionizing sources with overlapping HII regions is dealt with by assuming that the number of photons extracted from the radiation field of a given source by a given particle depends on the fraction of the total flux received by the particle supplied by that source. Ionised particles that are deprived of photons are allowed to recombine on their individual recombination timescales. If their ionization fractions fall below 0.1, they are considered to be fully neutral once more and they are then allowed to descend the cooling curve from \cite{1993A&A...273..318S}. Ionizing fluxes are assigned to stars and clusters in the same fashion as in Paper I and stellar sources with masses below 20 M$_{\odot}$ are again neglected. Owing to the generally smaller numbers of objects formed by these simulations, ionization is initiated in all cases when three objects are sufficiently massive to possess ionizing fluxes. We again continue the simulations after ionization is enabled for as close as possible to 3Myr, which we refer to as $t_{\rm SN}$, the time after which supernovae can be expected to begin detonating.\\
\indent Our model clouds initially have a Gaussian three--dimensional density profile. We seed the gas with a Kolmogorov turbulent velocity field whose total kinetic energy is equal in magnitude to 2.3 times the cloud's initial gravitational binding energy, so that the clouds are formally unbound. However, the convergent flows and shocks nevertheless rapidly generate gravitationally--unstable structures.\\
\section{Embedded cluster parameter space}
\begin{figure*}
     \centering
     \subfloat[Run UB]{\includegraphics[width=0.30\textwidth]{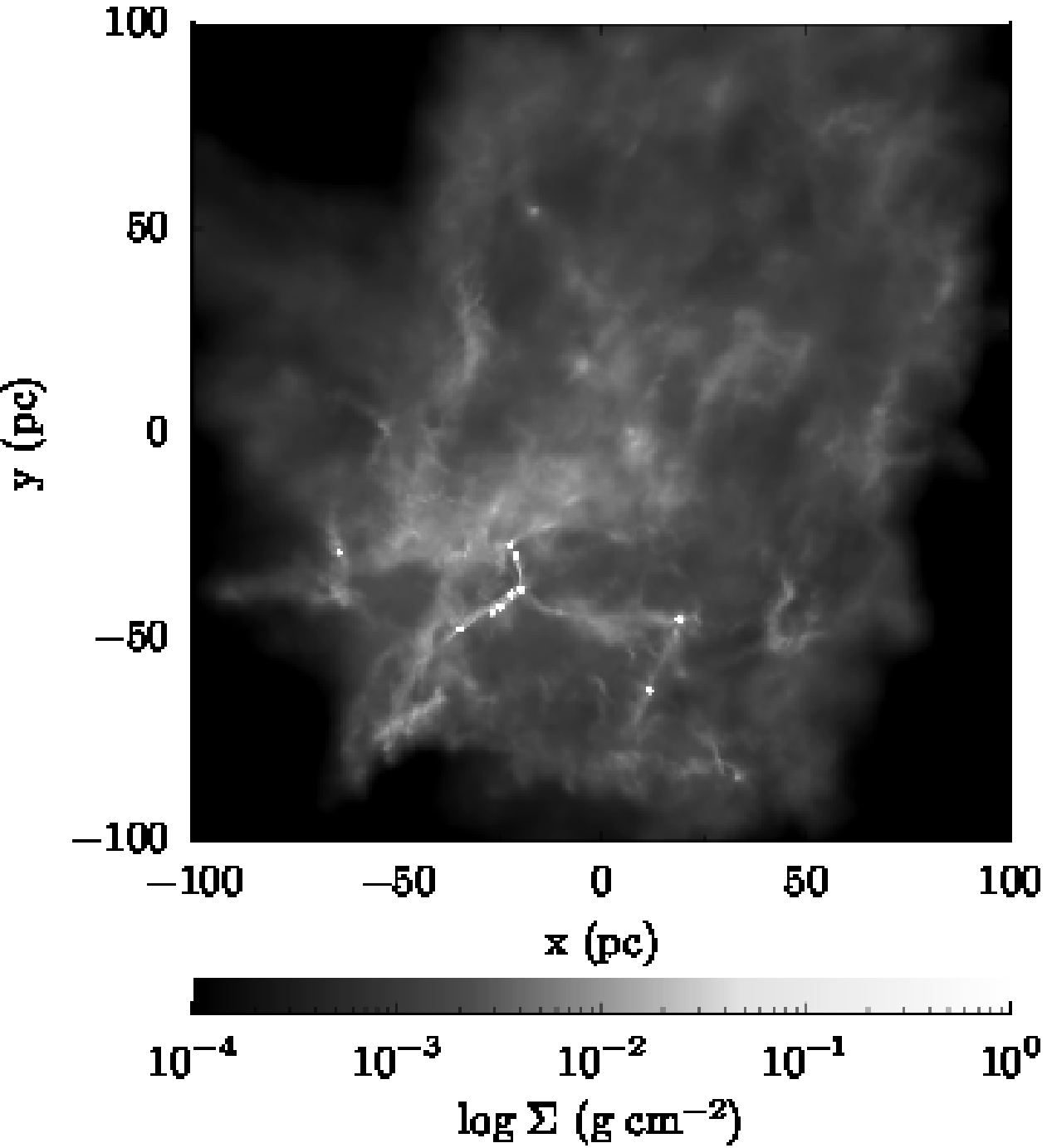}}     
     \hspace{.1in}
     \subfloat[Run UC]{\includegraphics[width=0.30\textwidth]{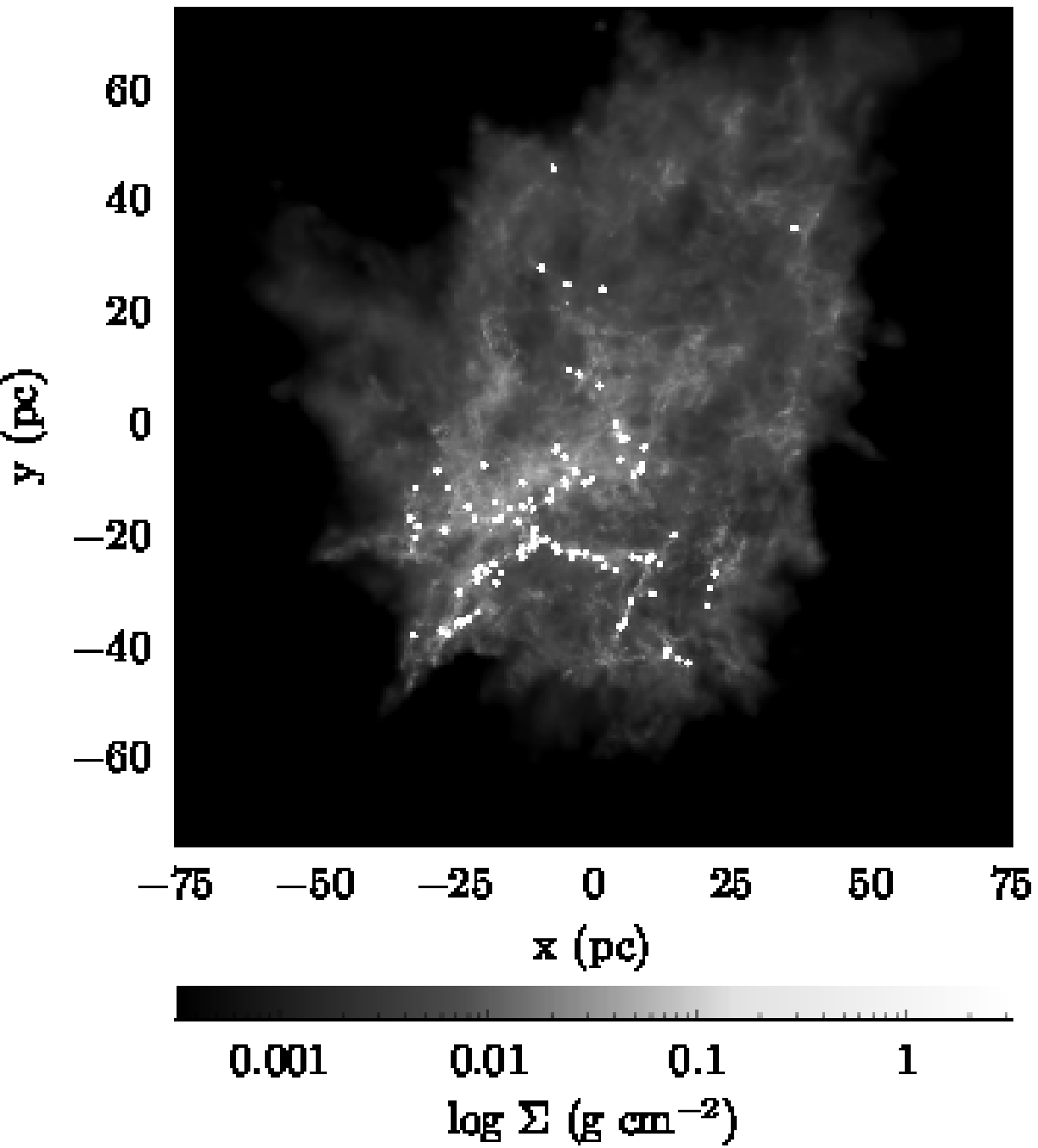}}
     \hspace{.1in}
     \subfloat[Run UZ]{\includegraphics[width=0.30\textwidth]{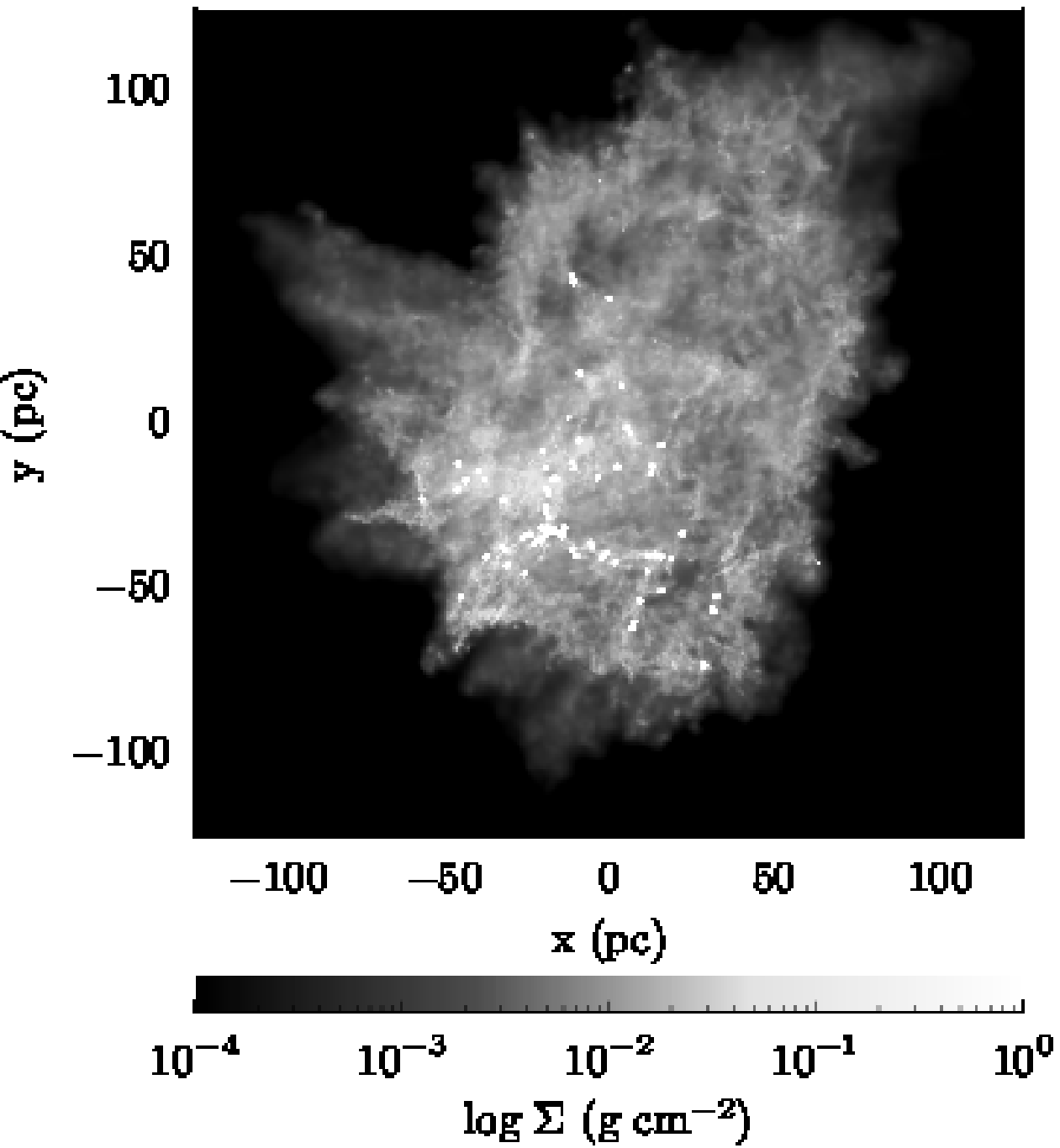}}
          \vspace{.1in}
     \subfloat[Run UU]{\includegraphics[width=0.30\textwidth]{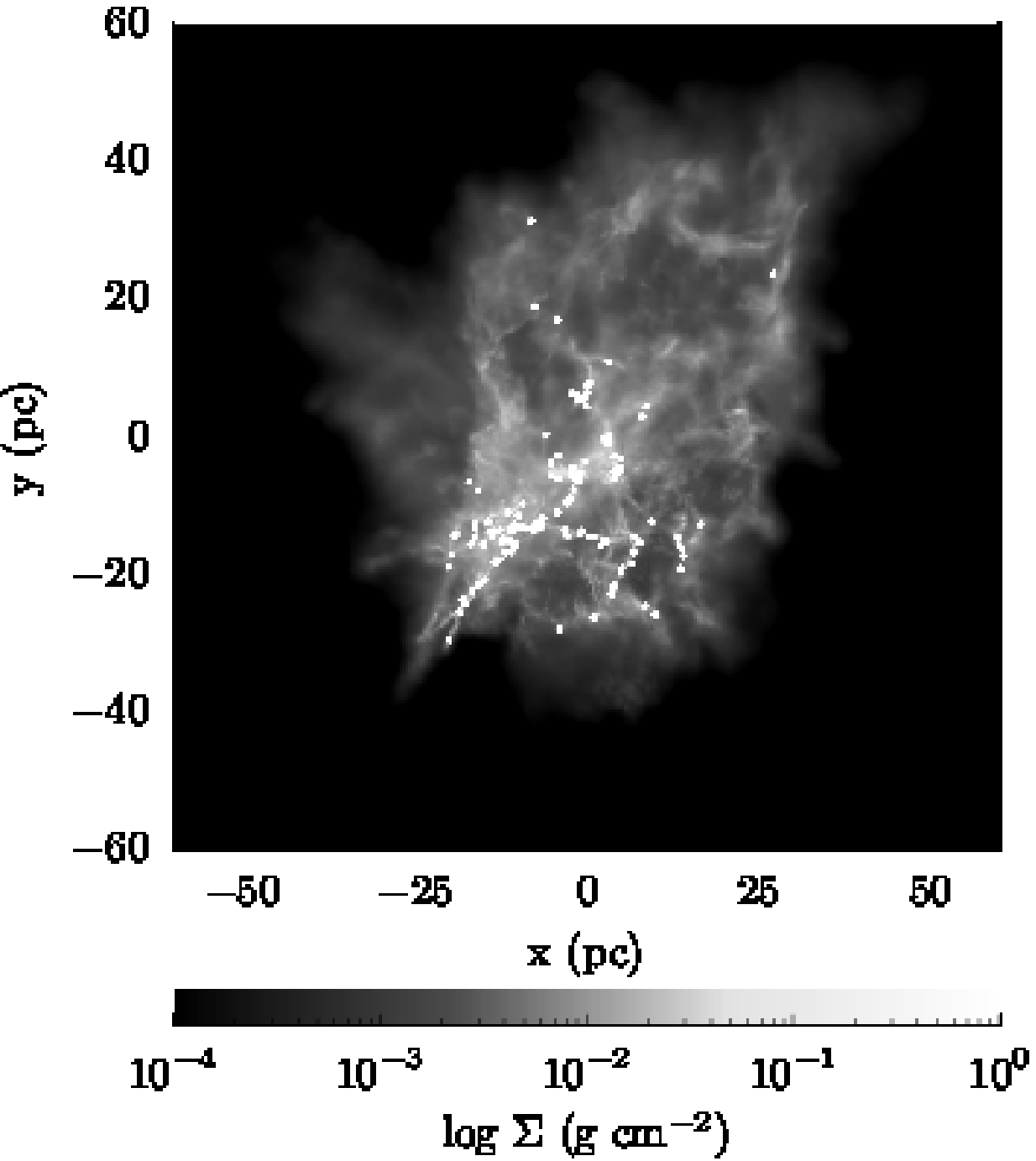}}
     \hspace{.1in}
     \subfloat[Run UV]{\includegraphics[width=0.30\textwidth]{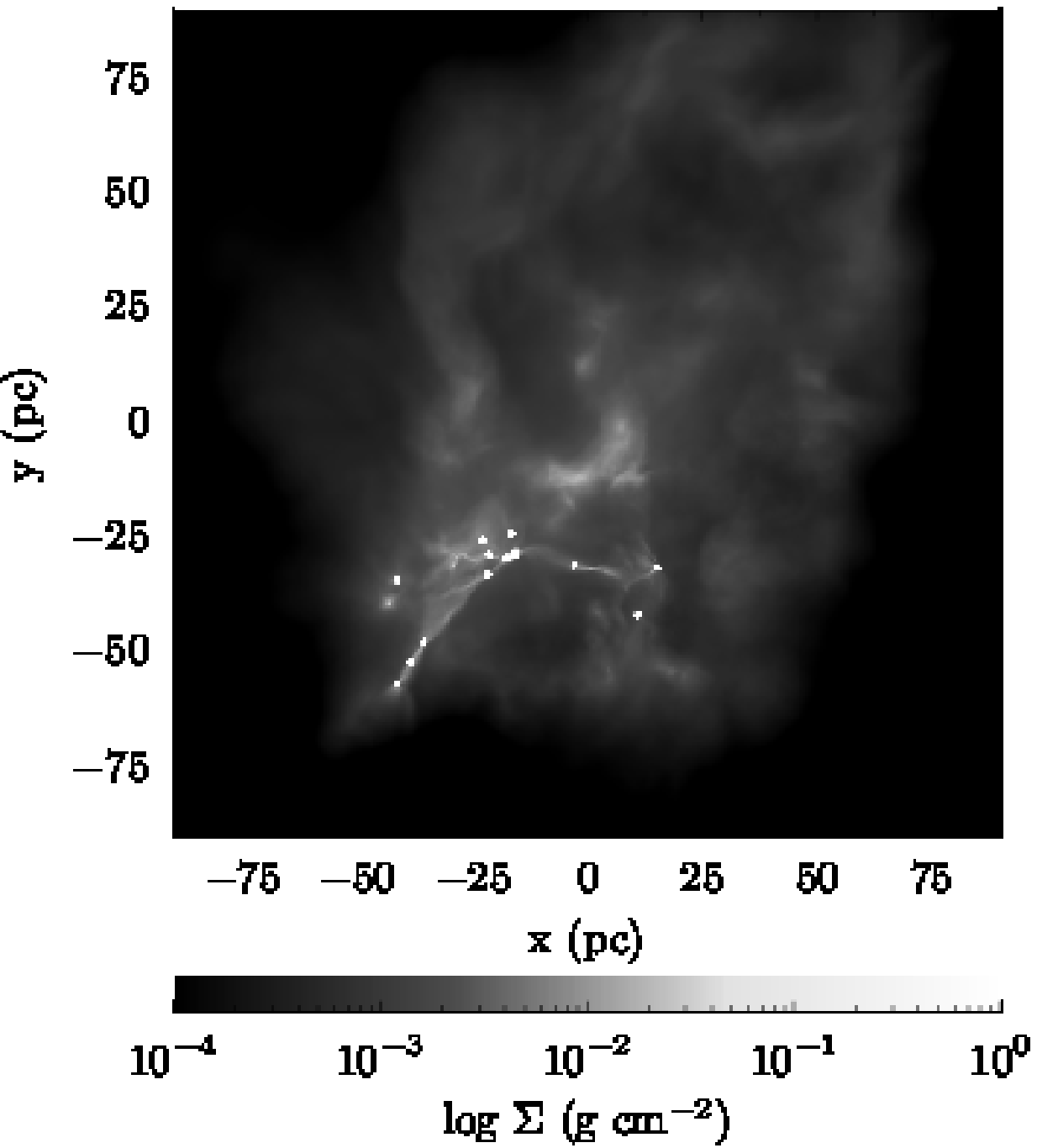}}
          \hspace{.1in}
     \subfloat[Run UF]{\includegraphics[width=0.30\textwidth]{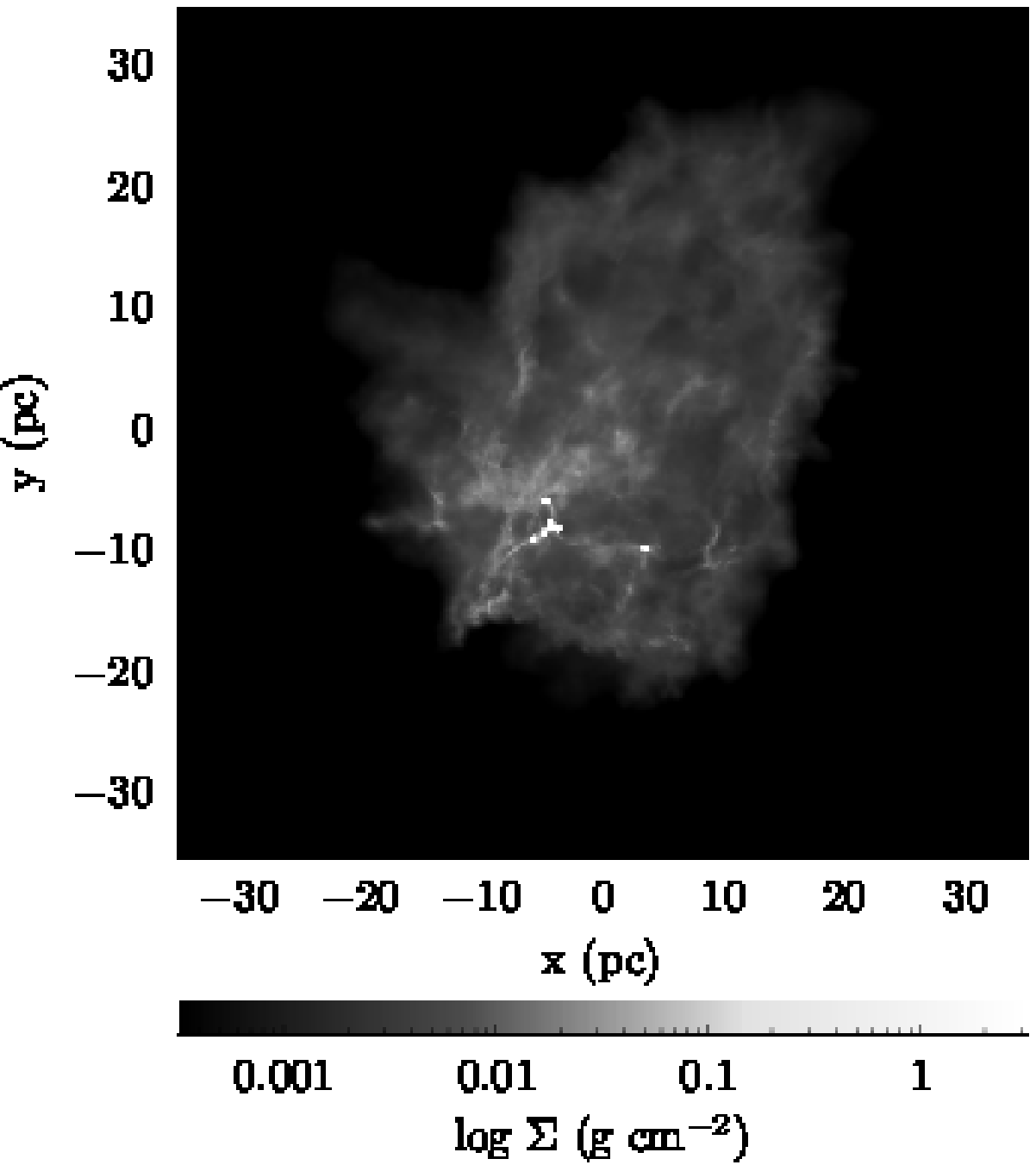}}
          \vspace{.1in}
     \subfloat[Run UP]{\includegraphics[width=0.30\textwidth]{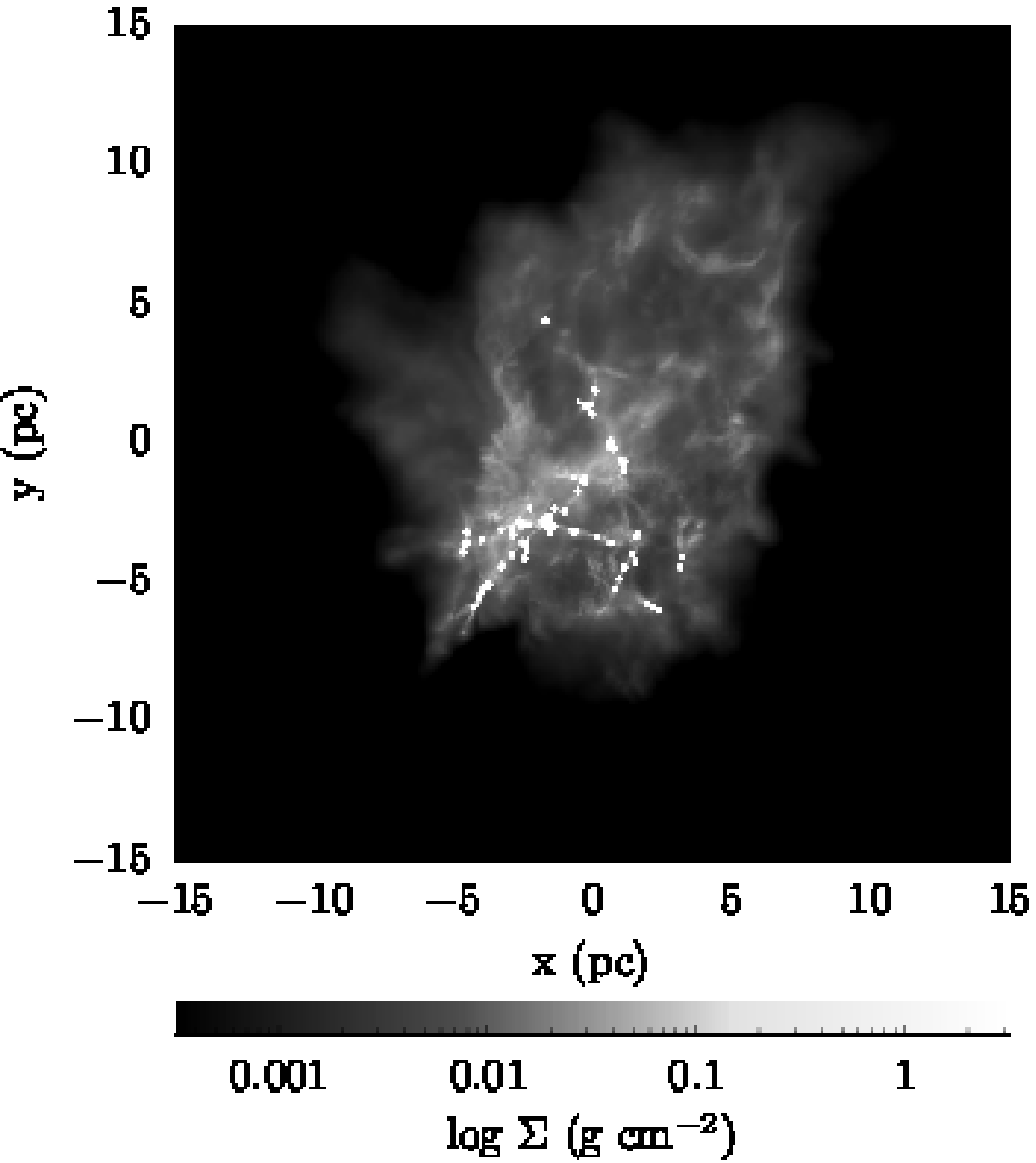}}
          \hspace{.1in}
     \subfloat[Run UQ]{\includegraphics[width=0.30\textwidth]{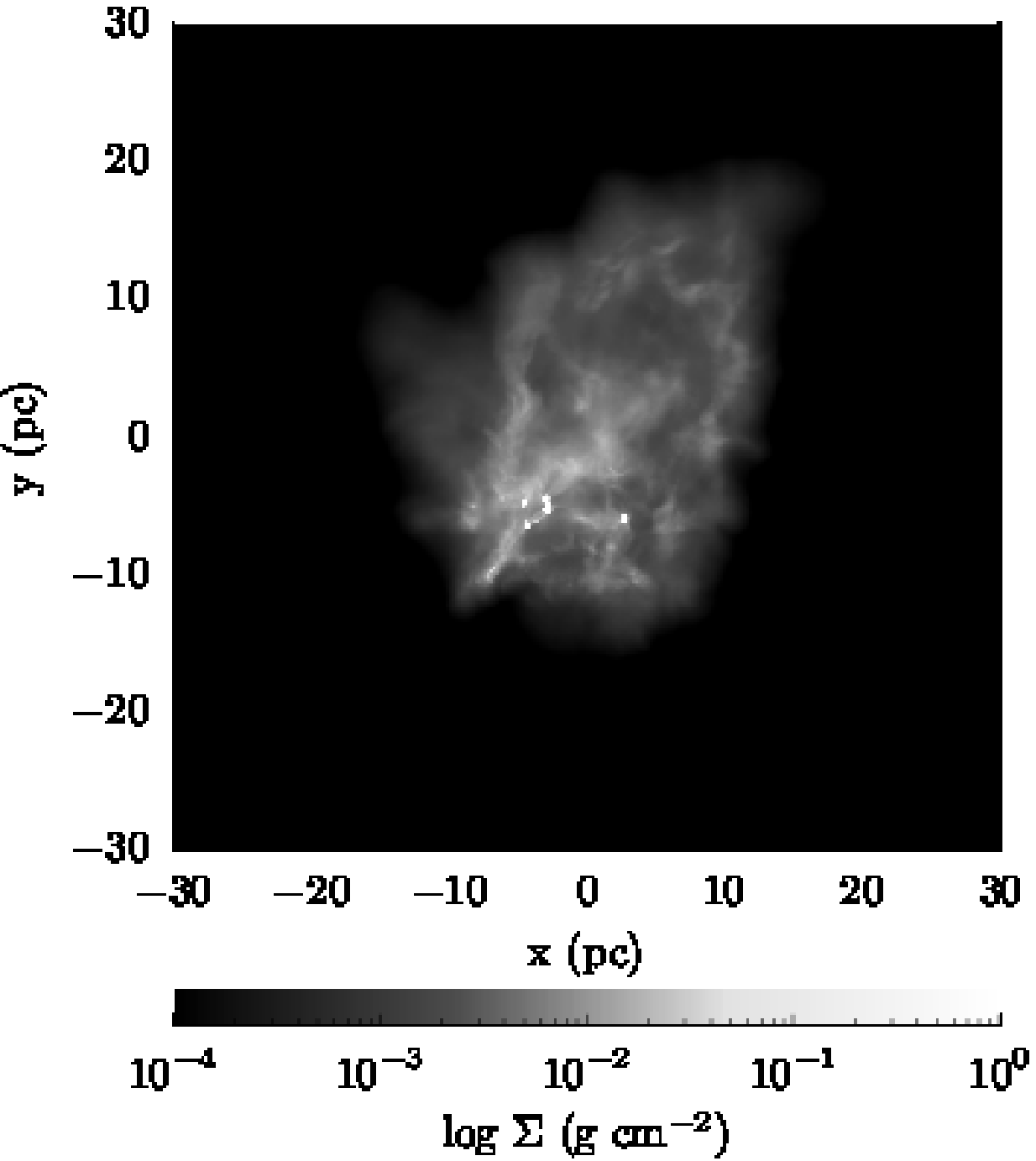}}
     \caption{Gallery of states of clusters at the times ionization begins in each case, as shown by column density maps observed down the $z$--axis. White dots represent sink particles (individual stars in runs UP, UQ and UF, clusters otherwise) and are not to scale. Note the different physical sizes and the different column density scales.}
   \label{fig:gallery_initial}
\end{figure*}

The initial conditions for the simulations presented here are based again on the work of \cite{2009ApJ...699.1092H} and are deliberately made very similar to those from Paper I. The clouds studied here are in fact rescaled copies of clouds from Paper I. All clouds in Paper I had an initial virial ratio of 0.7. All clouds in this work have an initial virial ratio of 2.3. The clouds are therefore all formally unbound at the beginning of the simulations, in the sense that their total turbulent kinetic energies are more than twice the magnitude of their gravitational potential energies. However, this does not imply that all the of the gas in the clouds is initially unbound, even formally. In fact, in all cases, $\approx35\%$ of the gas content of each cloud is unbound (i.e. has positive total energy in the centre--of--mass frame at the beginning of the simulation). Even though the virial ratio of the clouds is comfortably in excess of unity, almost two thirds of the gas is then initially bound and the clouds can be expected to form stars.\\
\indent The clouds studied in Paper I all had initial virial ratios of 0.7, so that their radii and therefore their mean densities and escape velocities remained roughly constant as they evolved. This is not true of the clouds studied here -- their radii are expected to increase and their mean densities and escape velocities therefore to decrease as they form stars. We illustrate this with an example from run UF in Figure \ref{fig:uf_rv} in which we plot the evolution with time of the 50, 75 and 90$\%$ Lagrange radii (solid lines), the velocity dispersion computed with these radii and the escape velocities computed at these radii (i.e. $v_{\rm esc}(r)=\sqrt{(2GM(r)/r)}$) up to the point when ionizing radiation is enabled.\\
\indent \cite{1998PhRvL..80.2754M} and \cite{1998ApJ...508L..99S} showed that the kinetic energy of an undriven turbulent velocity field decays with time approximately as $t^{-1}$ (whether the gas is magnetized or not) and that the velocity dispersion of the field therefore declines as $t^{-1/2}$. Figure \ref{fig:uf_rv} confirms that this is the case for the clouds considered here. A least--squares fit to the evolution of the kinetic energy within the 90$\%$ Lagrange radius reveals that the energy drops as $t^{-0.83}$ (the corresponding exponent for the 75$\%$ Lagrange radius is -1.08). During the epoch before ionization is enabled, the turbulent velocity dispersion within the 90$\%$ Lagrange radius falls by a factor of approximately two and the turbulent kinetic energy within this volume hence drops by a factor of four. The 90$\%$ Lagrange radius itself grows by a factor of approximately two and the cloud's gravitational potential energy decreases (in the sense of approaching zero) by the same factor. About one third of the loss of in kinetic energy is thus due to work done against the cloud's gravitational filed, while the remainder is due to decay of the velocity field. Note that, although the cloud is losing kinetic energy due to two processes -- turbulent decay and expansion against gravity -- since the radius of the cloud is increasing linearly with time, both of these processes cause the cloud's kinetic energy to fall as approximately $t^{-1}$. We confirmed that the turbulent kinetic energy in these model clouds is declining approximately linearly with time by computing the exponent for the decay of turbulence in the bound Run J cloud from Paper I (which neither expands nor contracts significantly before the onset of ionization), obtaining a value of -1.06.\\
\indent The overall change in the cloud virial ratio is then a decrease of almost a factor of two to a value of 1.3, approaching the typical value of around unity for the clouds reported in \cite{2009ApJ...699.1092H}. In addition, owing to the expansion, the escape velocity decreases by $\sim30\%$ (corresponding to a factor of $1/\sqrt{2}$). These values are typical for the clouds considered here.\\
\indent The clouds are therefore formally unbound at the onset of ionization, despite the fact that they are active (if slowly) forming stars. As we noted above, although the cloud virial ratios are in excess of unity, only about one third of the clouds' masses are energetically unbound. The reason for this apparent paradox is the the turbulent kinetic energy is initially not uniformly or smoothly distributed and, as the density field responds to the turbulence, this quickly becomes true of the potential energy as well. The global virial ratio is then not sufficient to characterize the clouds' behaviour. In addition to the total quantities of kinetic and potential energy changing with time, the distribution of energy evolves as well, as shown in Figure \ref{fig:energy_dist}. Here we plot the cumulative distributions of kinetic (red) and gravitational (blue) energy in the initial conditions (solid lines) and at the time when ionization is enabled (dashed lines). The dissipation of kinetic energy is more efficient in the cloud interior, so this region becomes drained of turbulent support, and the potential energy becomes more centrally concentrated. This eventually leads to regions of the cloud becoming unsupported, collapsing, and forming stars, despite the background cloud still being unbound and still expanding.\\
\indent In Table \ref{tab:init}, we present the initial properties of the clouds, but this can only give a limited idea of the environment in which the ionizing sources in the clouds will find themselves. Since these clouds are globally unbound, star formation occurs by definition in regions of the cloud which are not well described by global quantities. In Figures \ref{fig:mr}, \ref{fig:msigma} we plot the radii and RMS velocity dispersions of the clouds in the \cite{2009ApJ...699.1092H} sample as functions of mass, with the properties of our model clouds overlaid as points with error bars. The points represent the average quantities and the error bars show the maximum and minimum values of the relevant quantity for each model cloud up to the point when ionization is initiated (the masses of the clouds are regarded as constant). Here, `radius' is taken to be each cloud's 90$\%$ Lagrange radius.\\
\indent In Figure \ref{fig:gallery_initial} we show a gallery of column--density images viewed along the z--axis of all clouds and their embedded clusters at the respective epochs when ionization becomes active. This figure may be compared with Figure 6 in Paper I, although care should be exercised to note the length-- and column--density scales used. We have plotted these figures and the corresponding post--feedback images in Figure \ref{fig:gallery_final} to the same physical scales. Since some of the clouds expand substantially during the $t_{\rm SN}$ time window, they are of necessity plotted at rather small sizes in Figure \ref{fig:gallery_initial}. The general appearance of the clouds simulated here is very similar to those studied in Paper I, consisting of denser filamentary structures embedded in diaphanous low--density background gas. Star formation is once again largely concentrated along the filaments and particularly at filament junctions. However, the distribution of stars is in general sparser here, owing to the partially unbound nature of the clouds and their consequent lower star formation efficiencies (see the next section).\\
\begin{table*}
\begin{tabular}{|l|l|l|l|l|l|l|l|l|l|}
Run&Mass (M$_{\odot}$)&R$_{0}$(pc)&$\langle n(H_{2})\rangle$ (cm$^{-3}$) & v$_{\rm RMS,0}$(km s$^{-1}$)&v$_{\rm RMS,i}$(km s$^{-1}$)&v$_{\rm esc,i}$(km s$^{-1}$)&$t_{\rm i}$ (Myr) &t$_{\rm ff,0}$ (Myr)\\
\hline
UZ&$10^{6}$&45&149&18.2&9.4&13.8&4.33&2.9\\
\hline
UB&$3\times10^{5}$&45&45&10.0&4.6&7.6&9.40&6.0\\
\hline
UC&$3\times10^{5}$&21&443&14.6&6.0&11.1&4.03&1.9\\
\hline
UV&$10^{5}$&21&148&12.2&3.9&6.4&10.44&3.3\\
\hline
UU&$10^{5}$&10&1371&8.4&5.8&9.3&3.73&1.1\\
\hline
UF&$3\times10^{4}$&10&410&6.7&3.5&5.1&3.28&2.0\\
\hline
UP&$10^{4}$&2.5&9096&7.6&3.6&5.9&1.83&0.4\\
\hline
UQ&$10^{4}$&5.0&1137&5.4&2.6&4.1&3.13&1.2\\
\hline
\end{tabular}
\caption{Initial properties of clouds listed in descending order by mass. Columns are the run name, cloud mass, initial radius, initial RMS turbulent velocity, RMS turbulent velocity at the time ionization becomes active, the escape velocity at the same epoch, the time at which ionization begins, and the initial cloud freefall time.}
\label{tab:init}
\end{table*}
\section{Results}
\begin{figure*}
     \centering
     \subfloat[Run UB]{\includegraphics[width=0.30\textwidth]{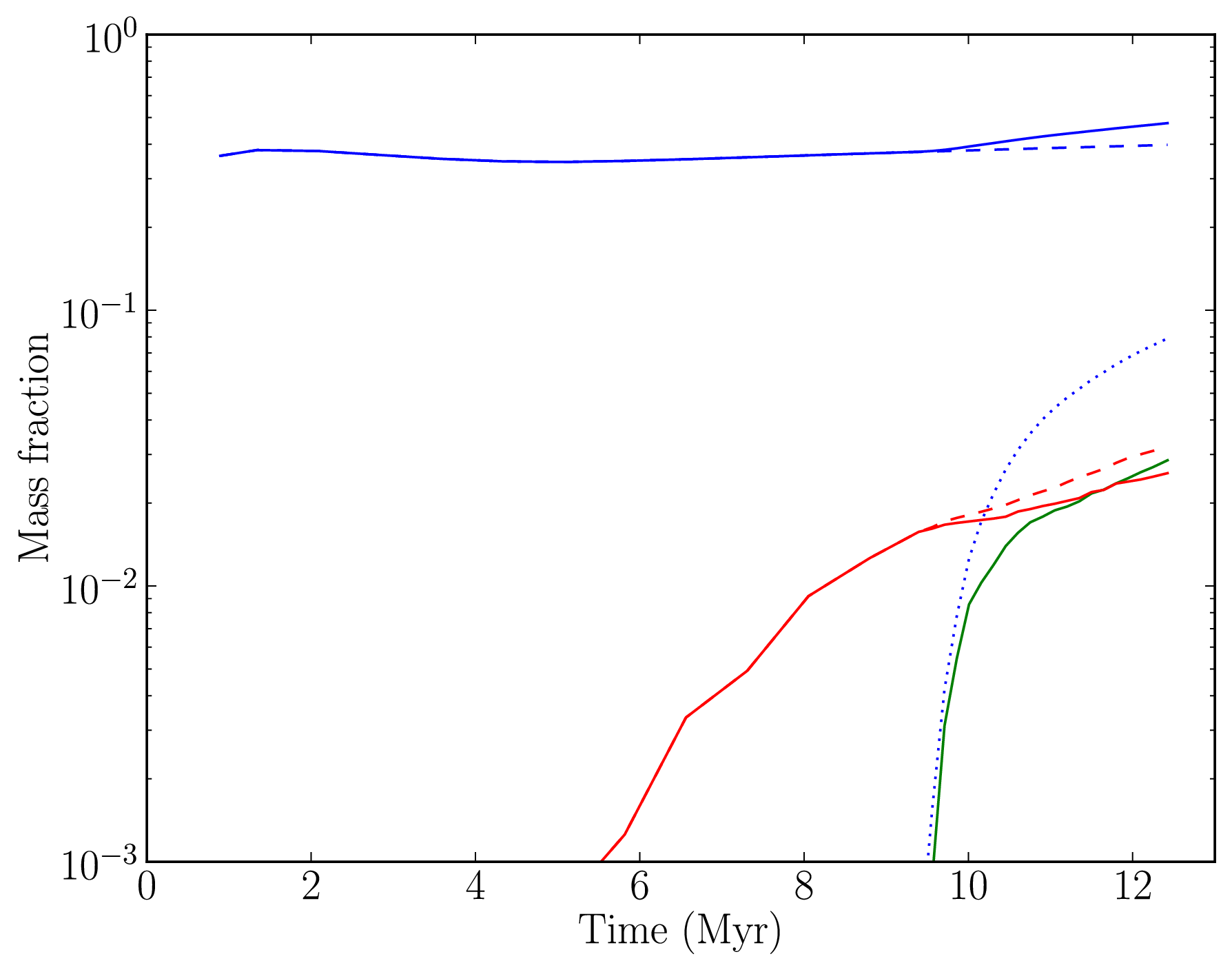}}     
     \hspace{.1in}
     \subfloat[Run UC]{\includegraphics[width=0.30\textwidth]{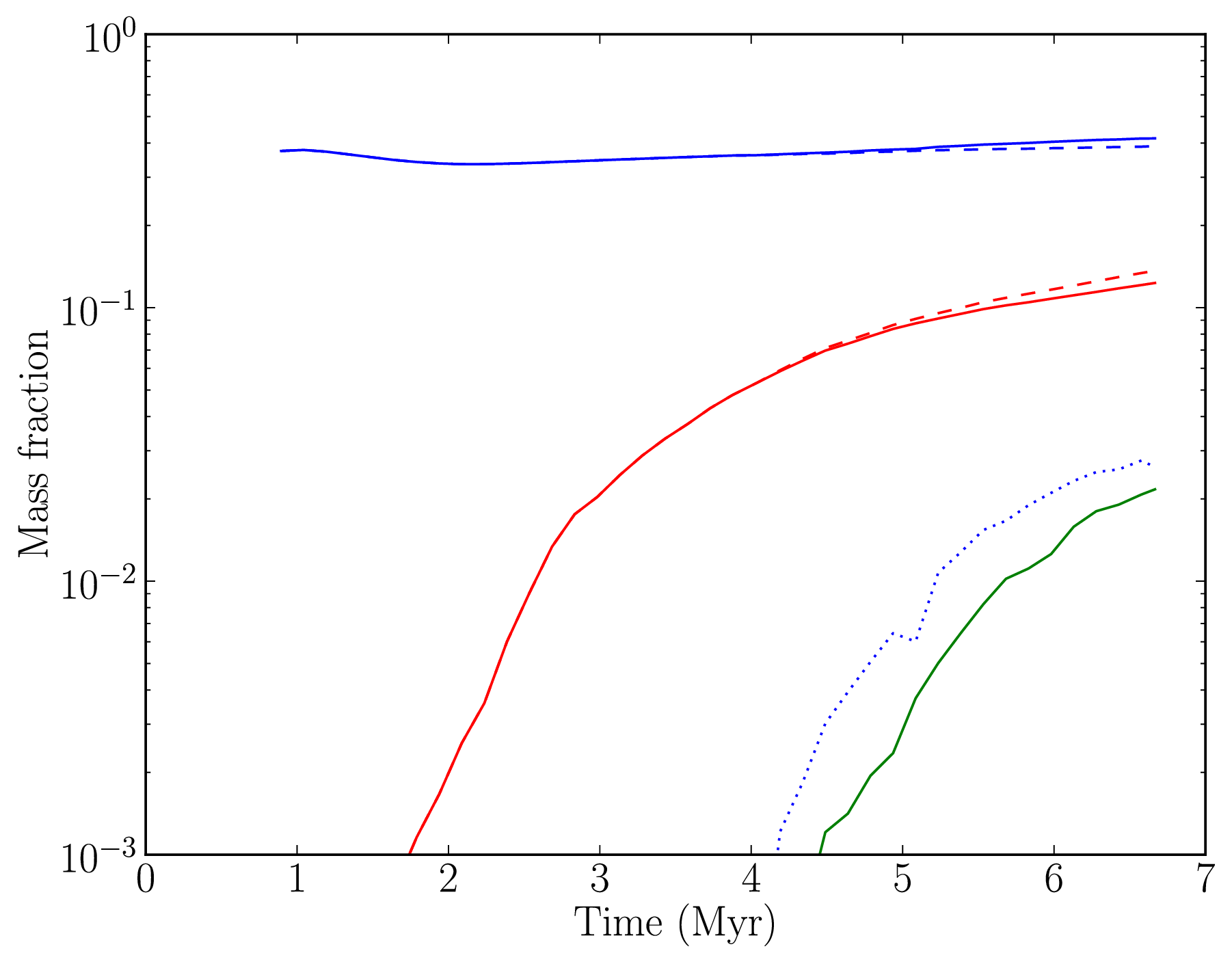}}
     \hspace{.1in}
     \subfloat[Run UZ]{\includegraphics[width=0.30\textwidth]{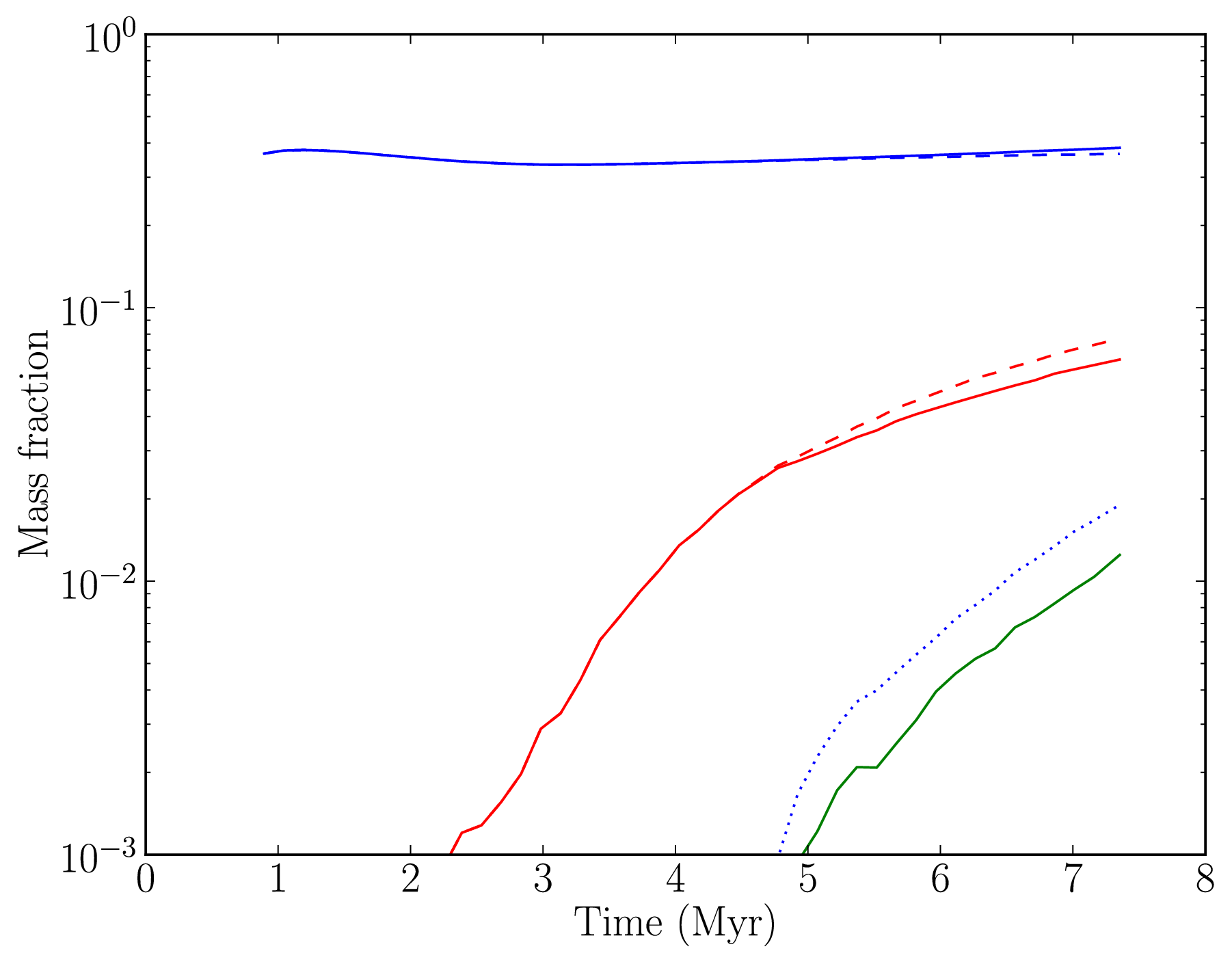}}
          \vspace{.1in}
     \subfloat[Run UU]{\includegraphics[width=0.30\textwidth]{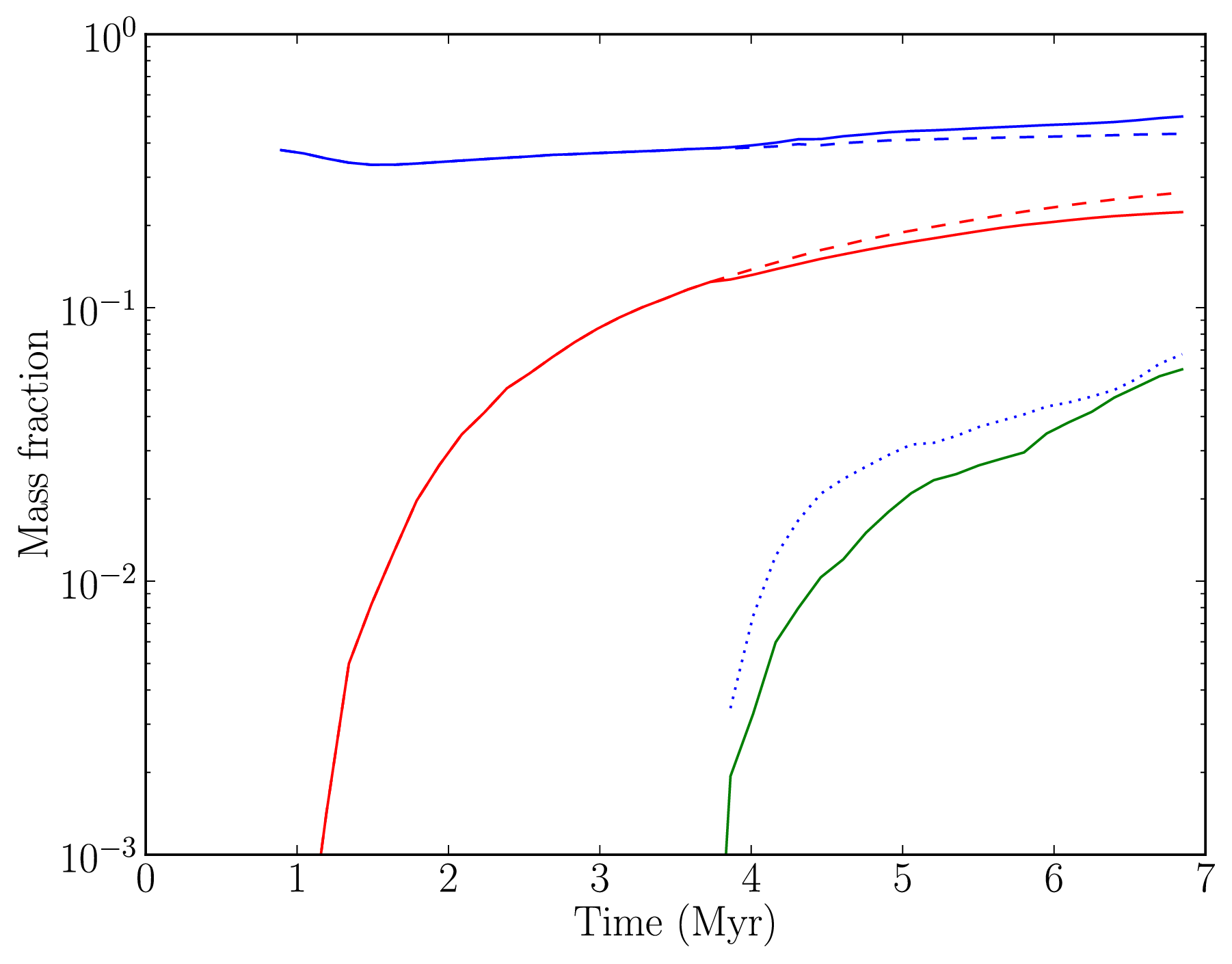}}
     \hspace{.1in}
     \subfloat[Run UV]{\includegraphics[width=0.30\textwidth]{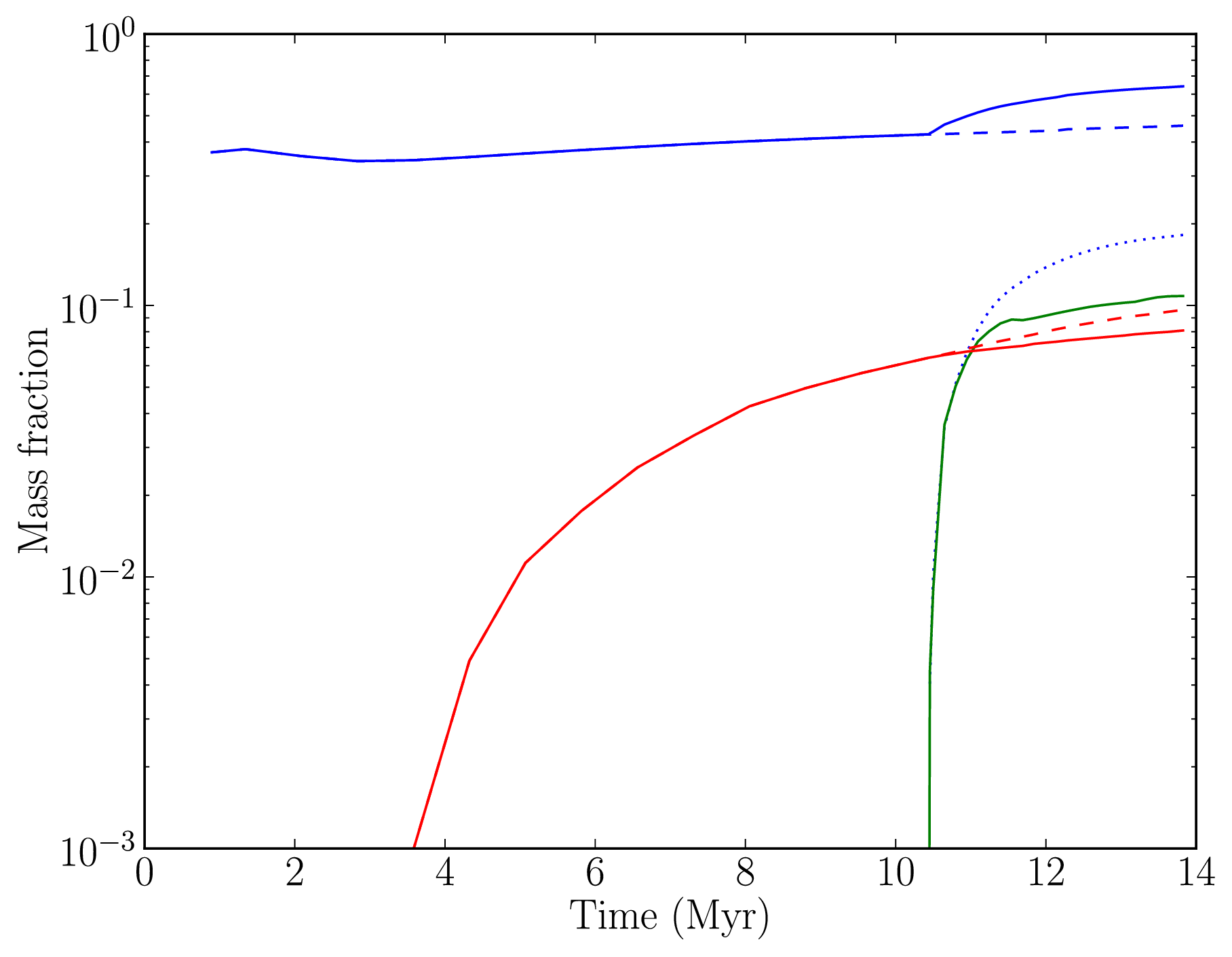}}
          \hspace{.1in}
     \subfloat[Run UF]{\includegraphics[width=0.30\textwidth]{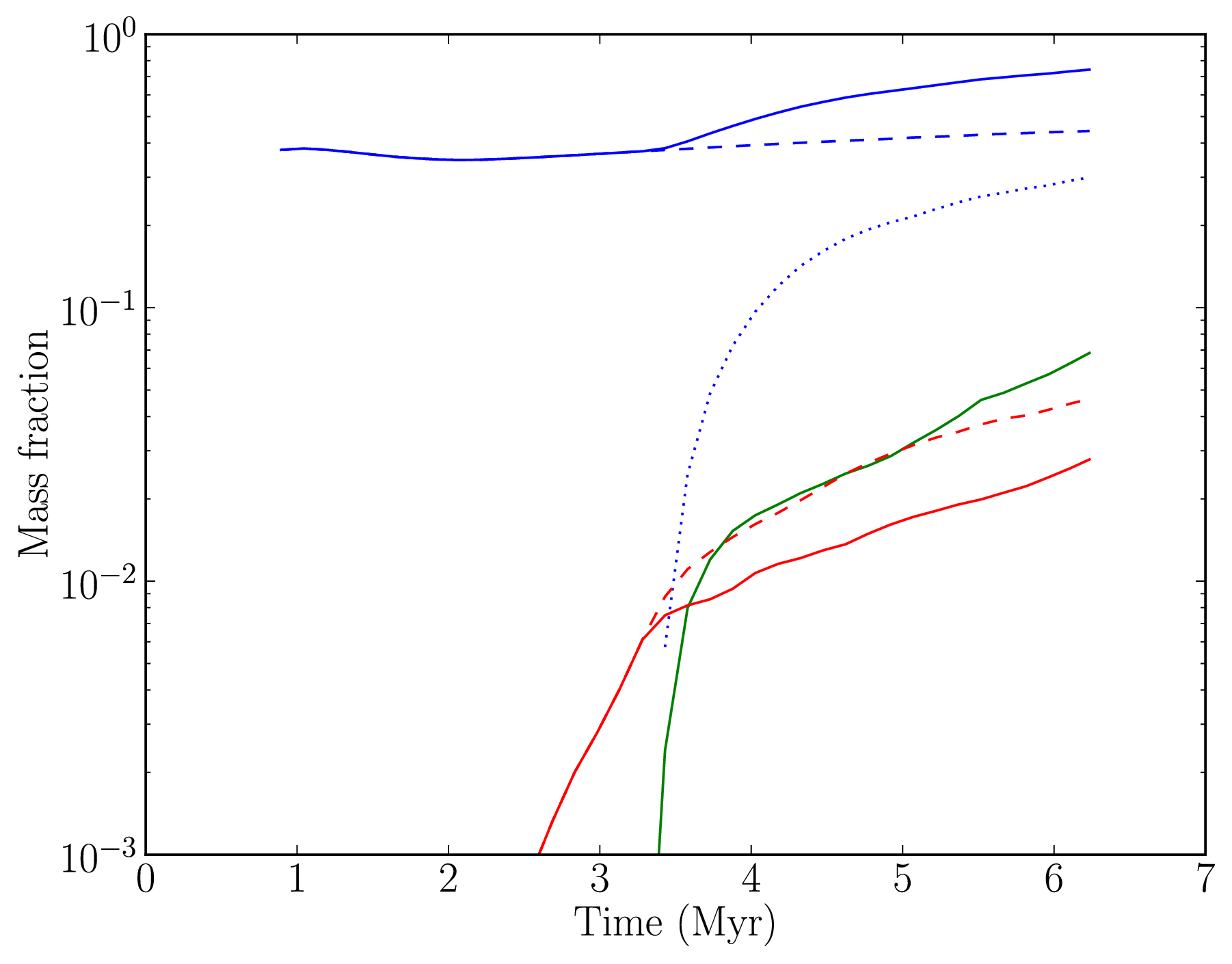}}
          \vspace{.1in}
     \subfloat[Run UP]{\includegraphics[width=0.30\textwidth]{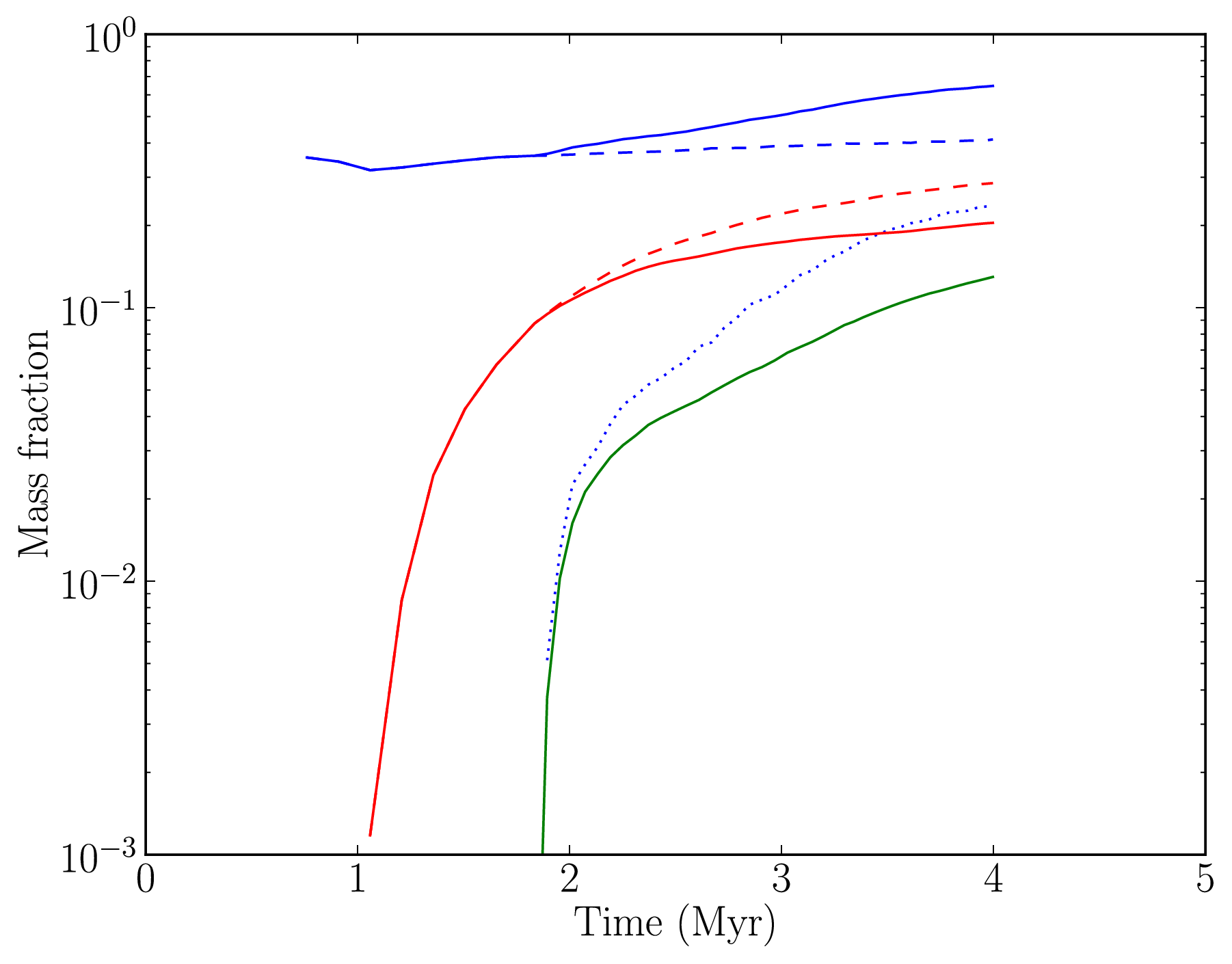}}
          \hspace{.1in}
     \subfloat[Run UQ]{\includegraphics[width=0.30\textwidth]{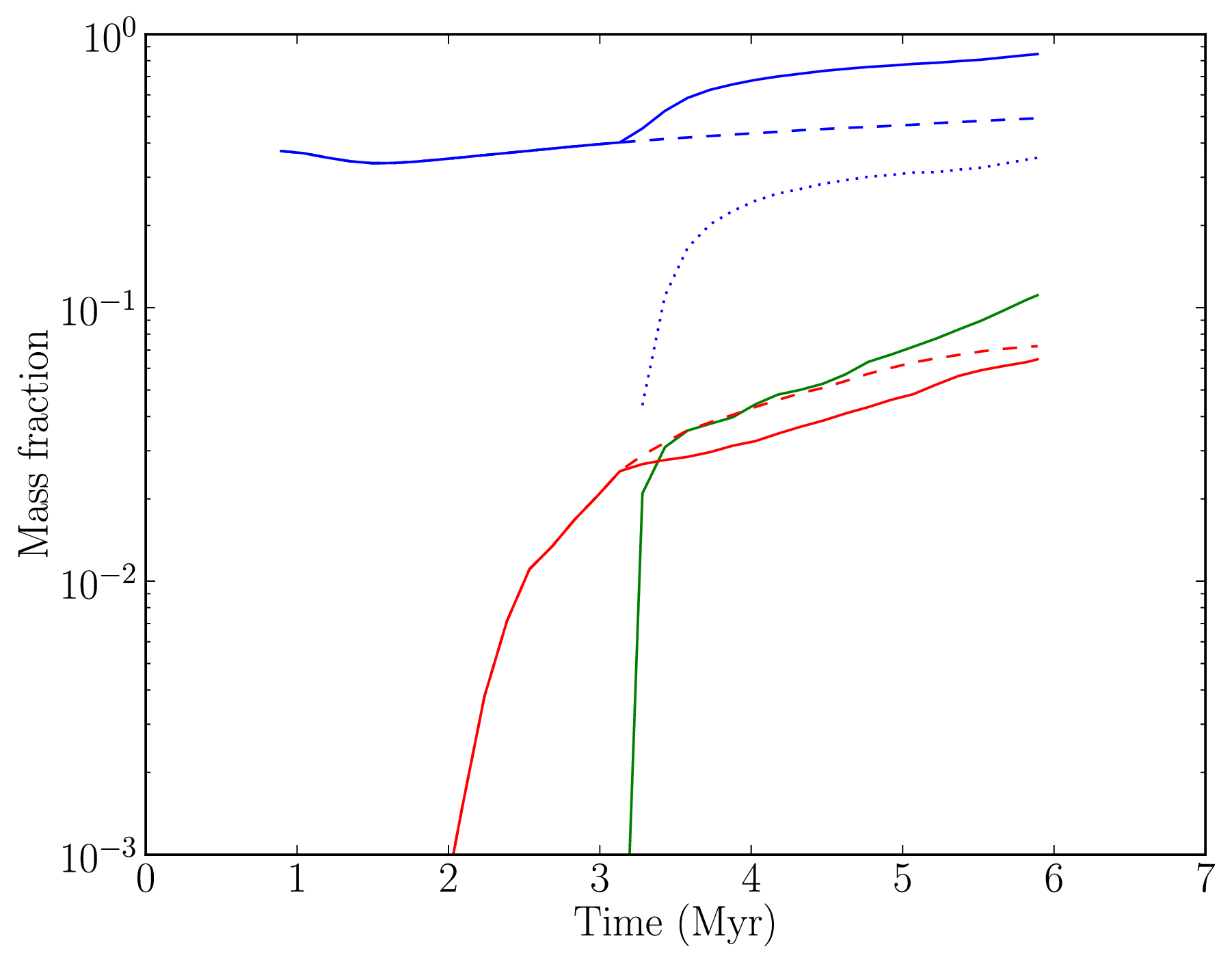}}
     \caption{Plots showing the evolution with time of the star formation efficiency (red), ionized gas fraction (green) and unbound mass fraction (blue) in all simulations. Solid lines are from feeback runs, dashed lines are from control runs. The dotted blue lines represent the unbound mass fraction in the ionized runs minus that in the corresponding control runs.}
   \label{fig:unbnd}
\end{figure*}

\begin{figure*}
     \centering
     \subfloat[Run UB]{\includegraphics[width=0.30\textwidth]{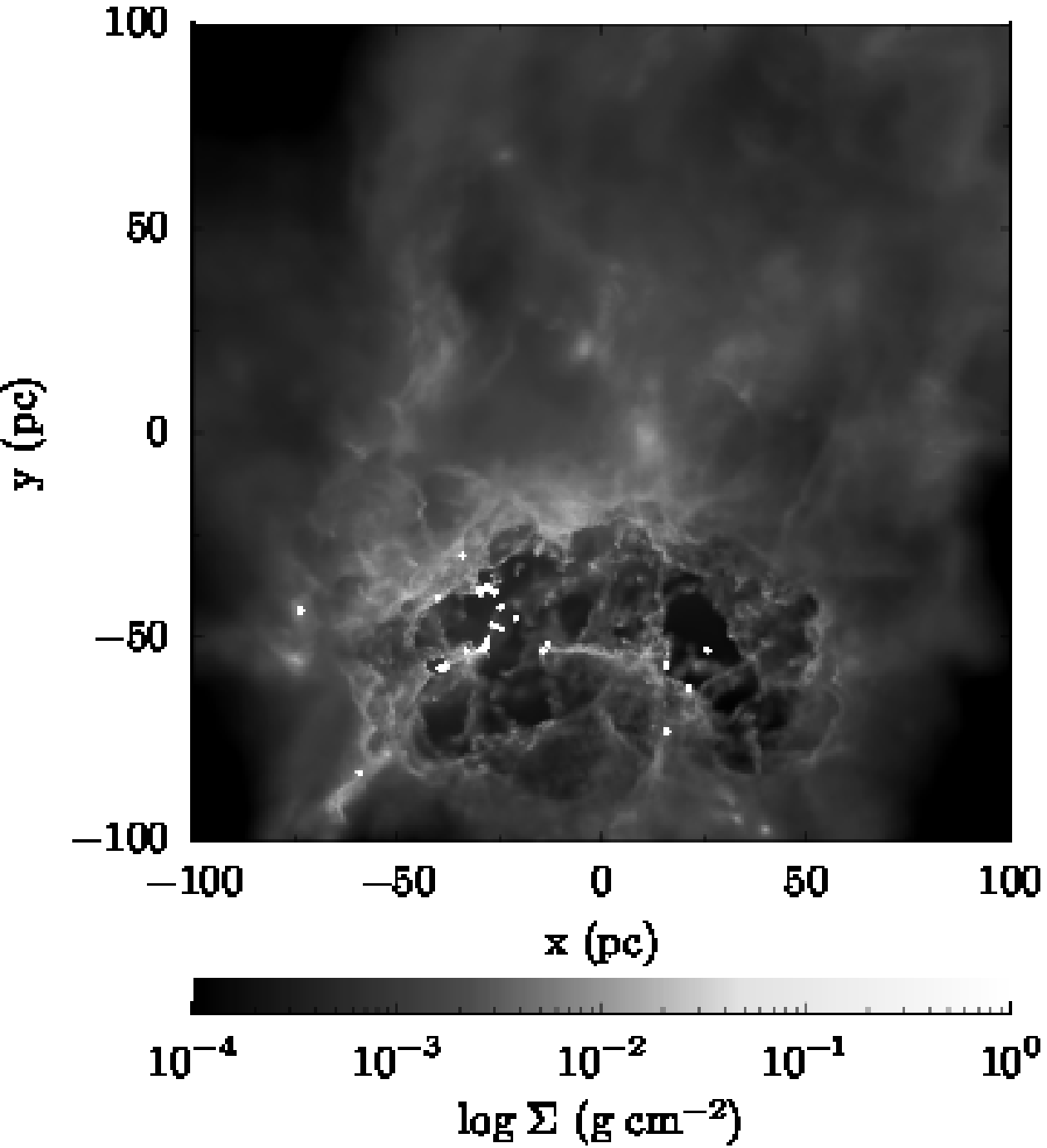}}     
     \hspace{.1in}
     \subfloat[Run UC]{\includegraphics[width=0.30\textwidth]{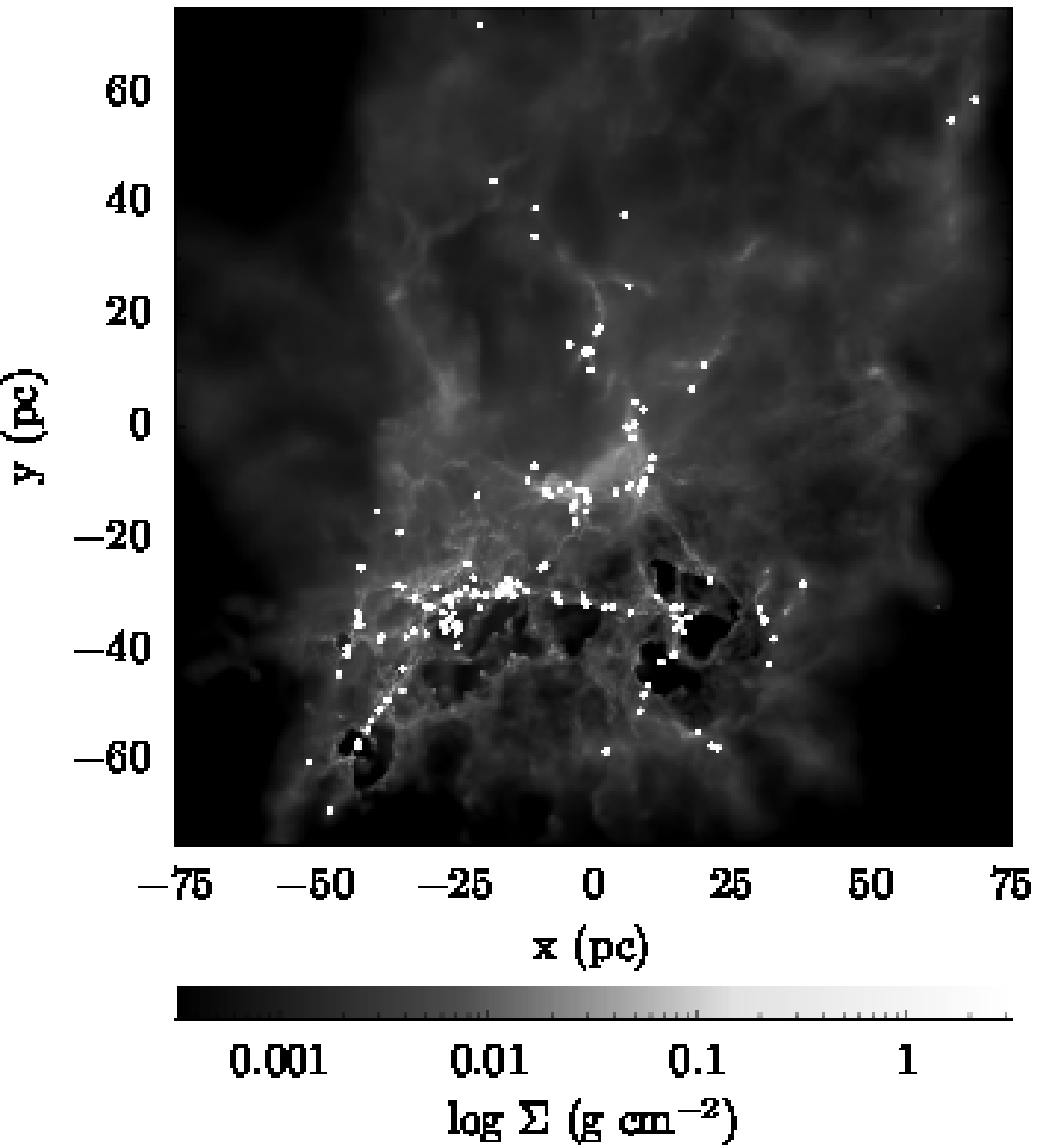}}
     \hspace{.1in}
     \subfloat[Run UZ]{\includegraphics[width=0.30\textwidth]{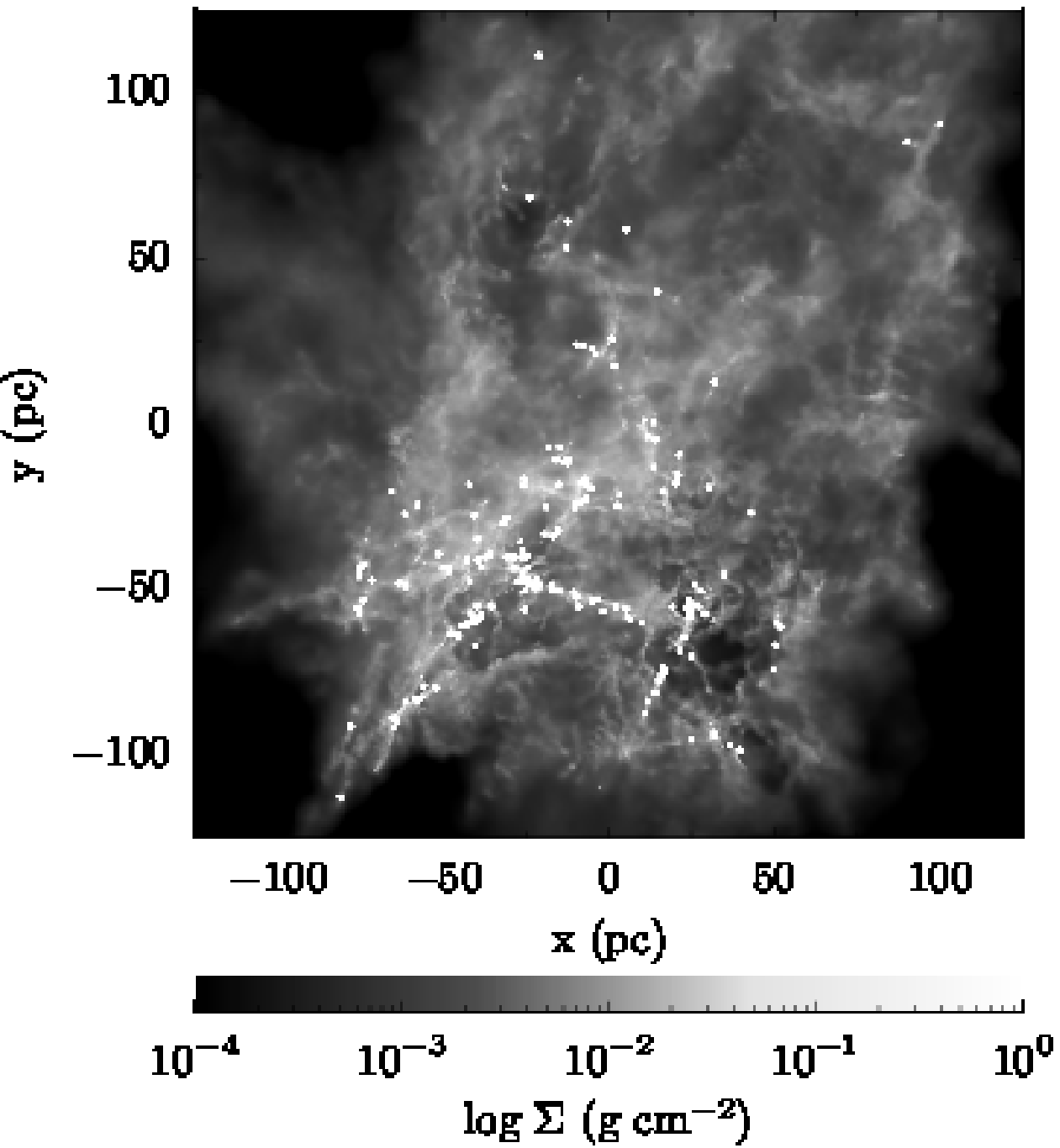}}
          \vspace{.1in}
     \subfloat[Run UU]{\includegraphics[width=0.30\textwidth]{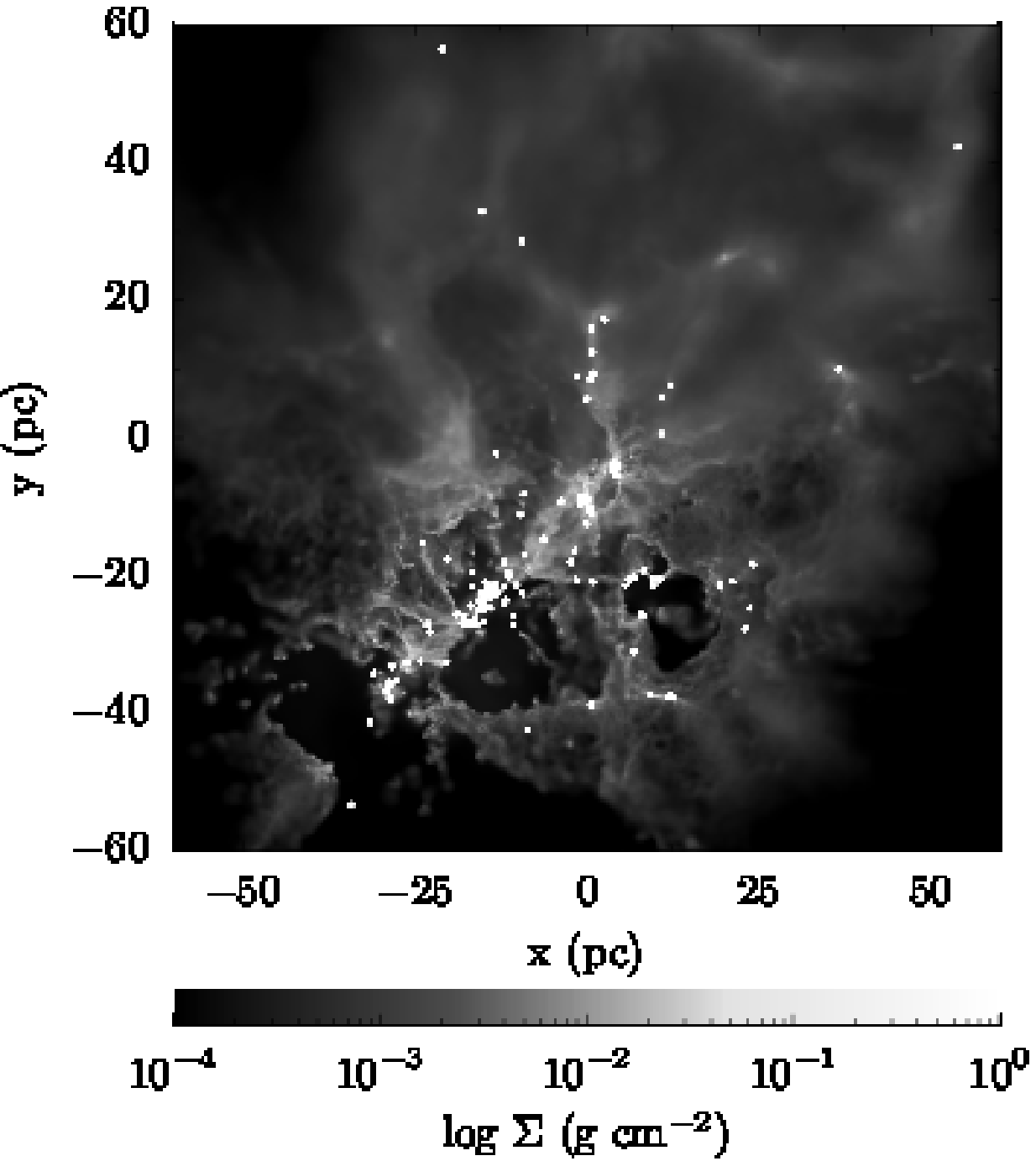}}
     \hspace{.1in}
     \subfloat[Run UV]{\includegraphics[width=0.30\textwidth]{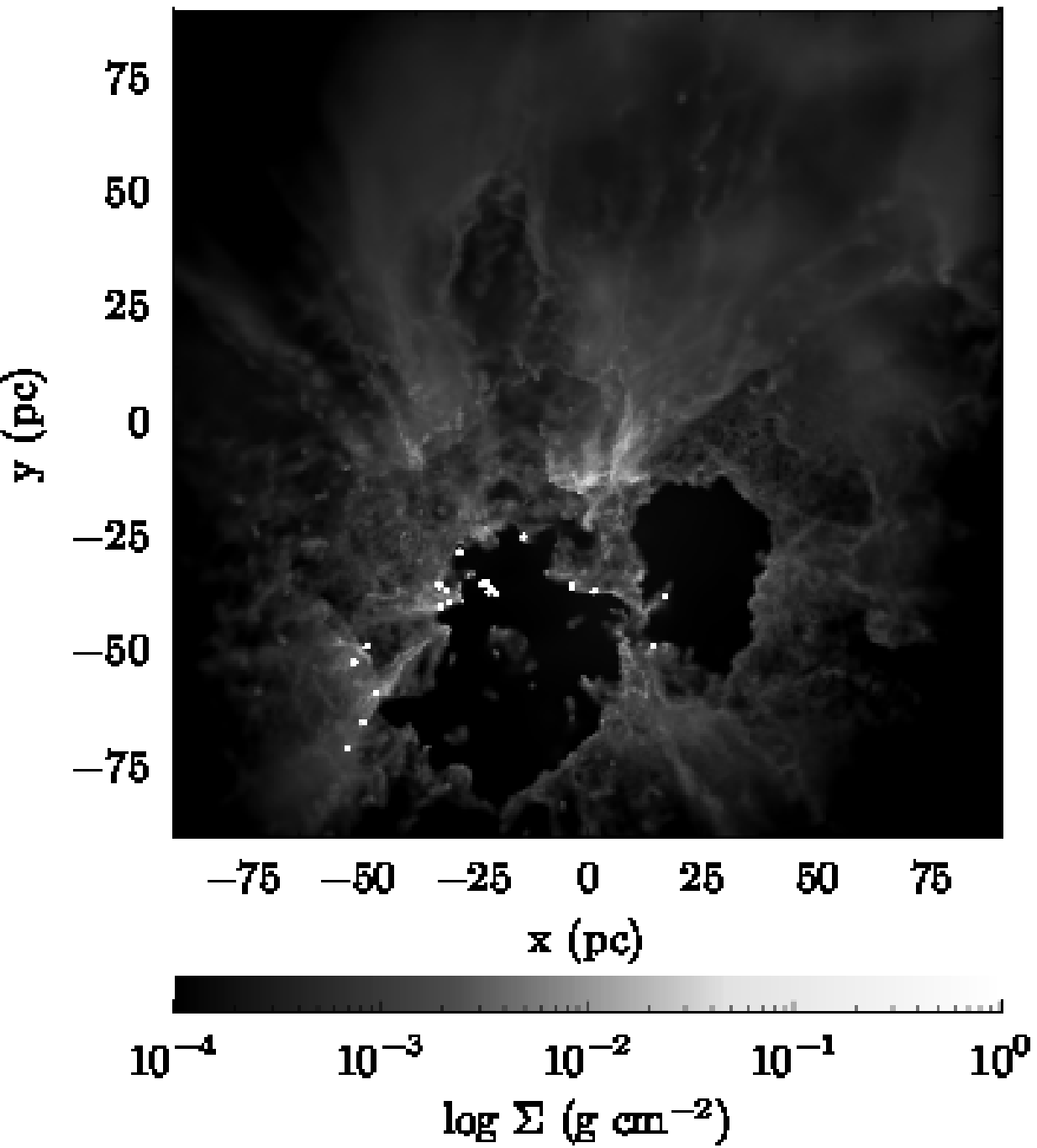}}
          \hspace{.1in}
     \subfloat[Run UF]{\includegraphics[width=0.30\textwidth]{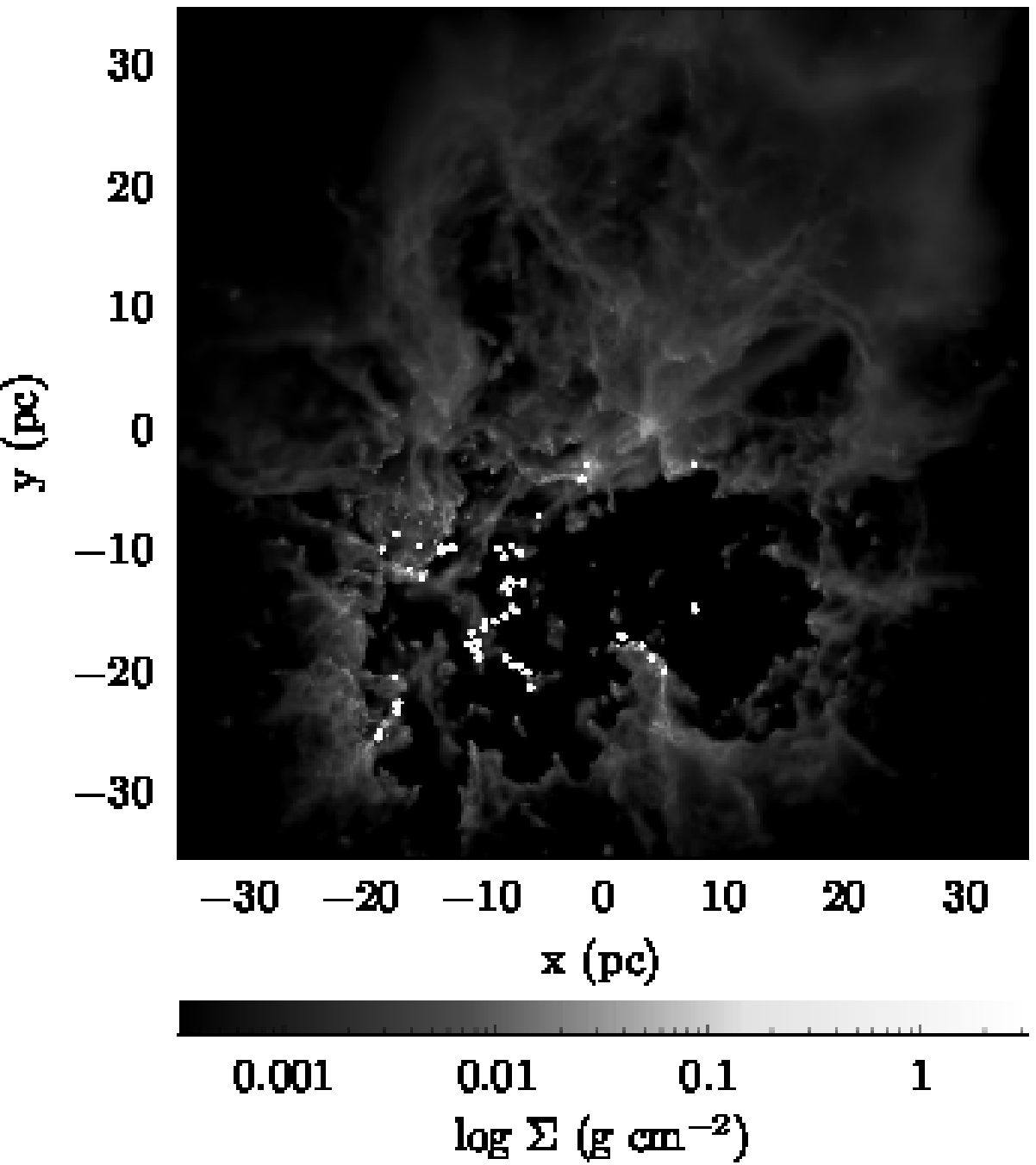}}
          \vspace{.1in}
     \subfloat[Run UP]{\includegraphics[width=0.30\textwidth]{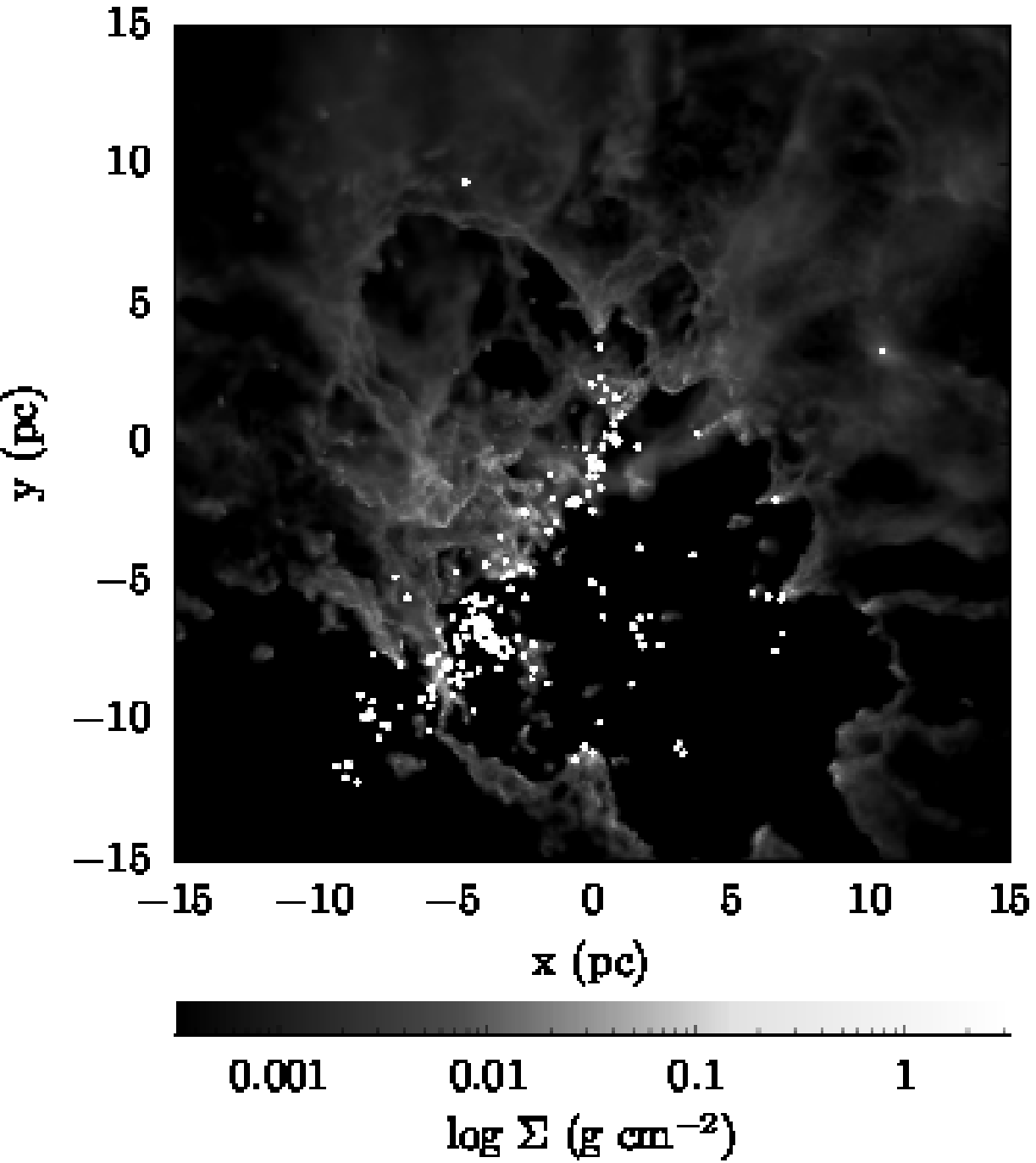}}
          \hspace{.1in}
     \subfloat[Run UQ]{\includegraphics[width=0.30\textwidth]{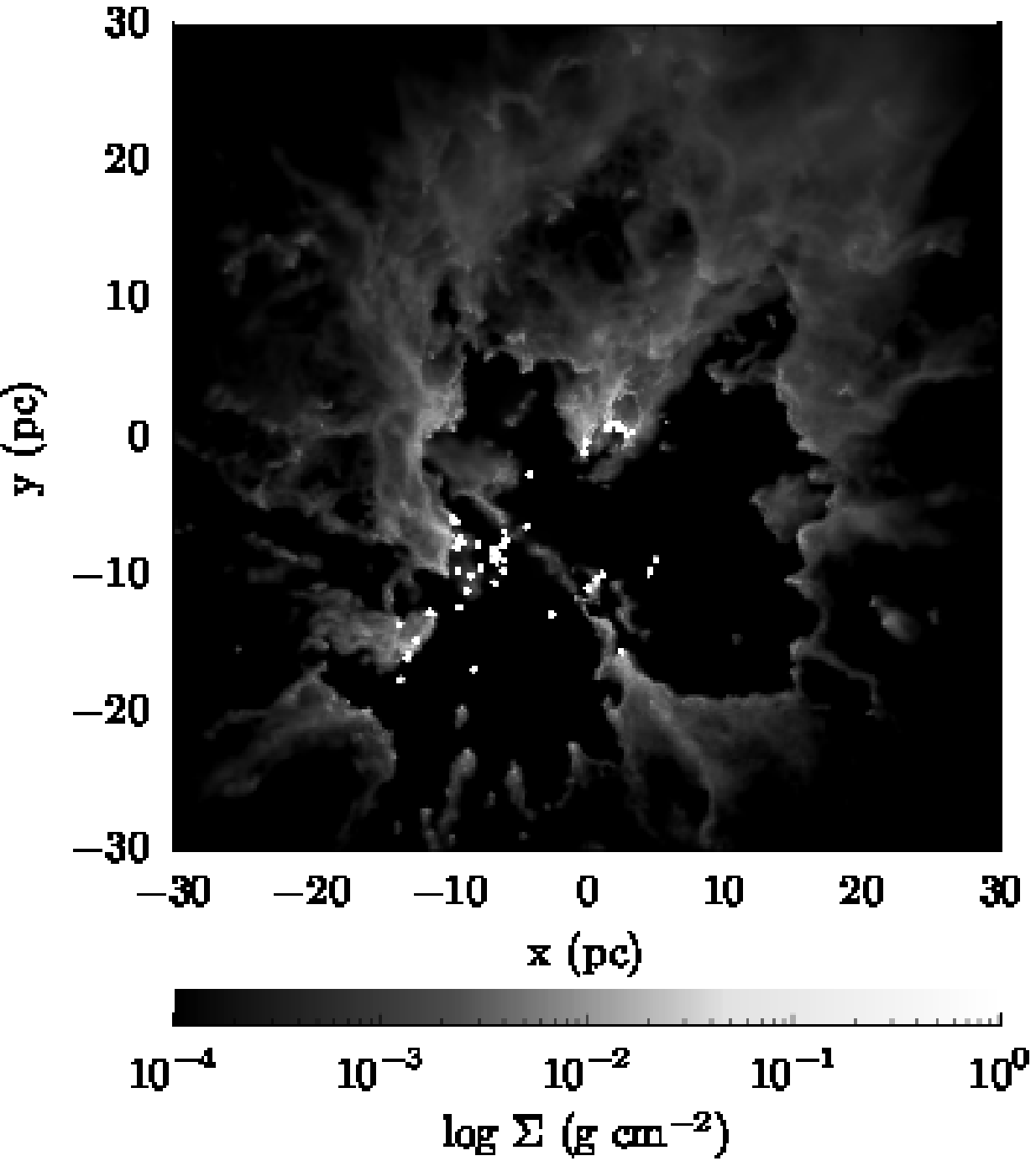}}
     \caption{Gallery of final states of clusters, as shown by column density maps observed down the $z$--axis. White dots represent sink particles (individual stars in Runs UP, UQ and UF, clusters otherwise) and are not to scale. Note the different physical sizes and the different column density scales.}
   \label{fig:gallery_final}
\end{figure*}

The principal aim of this study is to determine how effective ionization is in unbinding star--forming clouds which are already partially unbound thanks to their internal turbulent velocity fields. In Figure \ref{fig:unbnd} we plot the fraction of gas converted to stars (red lines), the fraction of gas ionized (green lines) and the fraction of gas unbound (blue lines) as functions of time in all our simulations. In order to isolate the effects of ionizing feedback directly, we plot in all cases the corresponding curves for control runs without feedback as dashed lines.\\
\indent All clouds are initially $\sim35\%$ unbound in the sense that this fraction of the gas has positive energy in the centre--of--mass frame when the simulations begin. As the simulations progress, the clouds expand and some of this unbound gas moves to large radii, making the cloud potential well shallower and thus unbinding more peripheral gas. In addition, our equation of state implicitly heats very low--density gas. In the calculations presented in Paper I, this effect was negligible on the timescales of interest in the clouds which were able to form stars, unbinding at most a few percent of the gas over the course of the simulations. The gas unbound in this way was the extremely tenuous material near the edges of the clouds, which effectively evaporated away. Since the clouds in Paper I were bound and therefore not expanding, the quantity of such very low--density material was always low. The clouds in the simulations presented here are expanding however and it is not obvious therefore that significant fractions of the clouds should not become of low enough density to be evaporated away.\\
\indent We see, however, from the dashed blue lines in Figure \ref{fig:unbnd} that this evaporative effect is again very slow and that the unbound gas fraction in simulations without feedback stays approximately constant. From this point of view, the ionizing sources in all calculations have a similar task ahead of them in attempting to unbind the clouds. To evaluate how much material is unbound solely by the action of photoionization, we subtract from the quantities of gas unbound in the ionized run at each time, the quantity of gas unbound at the same epoch in the corresponding control run to generate the dotted blue lines, which therefore represent the fraction of gas expelled by ionization alone as a function of time.\\
\indent It is immediately apparent from Figure \ref{fig:unbnd} that the unbound clouds exhibit a narrower range of star formation efficiencies ($\sim3$-$20\%$) than seen in Paper I ($\sim3$-$70\%$) and that, as expected \citep[e.g][]{2008MNRAS.386....3C}, the star formation efficiencies in the unbound clouds are generally lower either with or without feedback active, with most failing to achieve 10$\%$ efficiency over the durations of these simulations. It is also clear that, as reported in Paper I and Dale, Ercolano and Bonnell (2012b), in all cases feedback decreases the star formation efficiencies emerging at the ends of the simulations, although never by a very large factor, so that the efficiencies in the feedback--influenced clouds are still mostly a few to ten percent.\\ 
\indent As in the bound clouds studied in Paper I (with the exception of Run X), the clouds' global ionization fractions are all in the range of a few to ten percent for most of the simulation times. The dense gas, the accretion flows present and the large sizes of the clouds strongly restrict how much material the ionizing sources are able to ionize.\\
\indent The behaviour of the unbound clouds is also similar to that of the bound clouds in that the most massive clouds (UB, UC and UZ on the top row of Figure \ref{fig:unbnd}) are once again those least affected by feedback, with ionization unbinding $\lesssim10\%$ of additional gas on top of that already unbound at the onset of feedback.\\
\indent All clouds except Run UZ show evidence of bubble morphology which was much rarer in the clouds studied in Paper I, and in fact most of the clouds presented here have a strikingly similar appearance if the size scales are ignored, even though ionization has had very different dynamical effects on the clouds. Most of the clouds possess a few well--defined and generally well--cleared bubbles. The radii of the bubbles are virtually all of order 10pc. Since $c_{\rm HII}t_{\rm SN}\approx30$pc, this sets a maximum expected size for the expansion of confined bubbles (if their expansion is attributed solely to thermal overpressure, and not to continued ionization of substantial further volumes of the clouds), and the size observed here imply that the bubbles have expanded relatively freely into the clouds. This was only seen in a few cases -- Runs A, D and I -- in the bound--cloud simulations in Paper I, corresponding to the lowest density clouds in each mass group.\\
\indent It is not surprising that bubble formation by photoionization is easier in a low--density environment since the initial driving velocity of HII regions is always $c_{\rm HII}$ and mass--loading of the driven shell as a function of distance travelled is slower in low--density environments. Shells in lower density environments decelerate more gently and are thus able to travel further in the fixed time interval imposed by $t_{\rm SN}$. However, the differences in shell--driving in clouds of different densities are much less marked in the simulations presented here, implying that there must be some other effect at work. The above argument is simple and strictly only applies when the background cloud is stationary.\\
\indent In the turbulent clouds studied in this series of papers, the turbulent velocity field also likely plays a role. As was remarked earlier, the bound clouds from Paper I are all quite close to virial equilibrium and, if anything, are slowly contracting. At any given point in these clouds, the turbulent flows are travelling in random directions but, averaged over a sufficiently large volume, the flows have divergences close to zero or slightly negative. The unbound clouds presented here, by contrast, are all expanding, so that their turbulent velocity fields (again, when averaged over substantial volumes) have positive divergence. Since the expansion required to generate a bubble is a divergent flow, it is plainly easier to produce bubbles in a medium which already has a divergent velocity field, unless the background flow velocity is faster than the bubble expansion speed -- rarely the case here.\\
\section{Effects of photoionization on model clouds and clusters}
\subsection{Gas expulsion}
\begin{figure}
\includegraphics[width=0.45\textwidth]{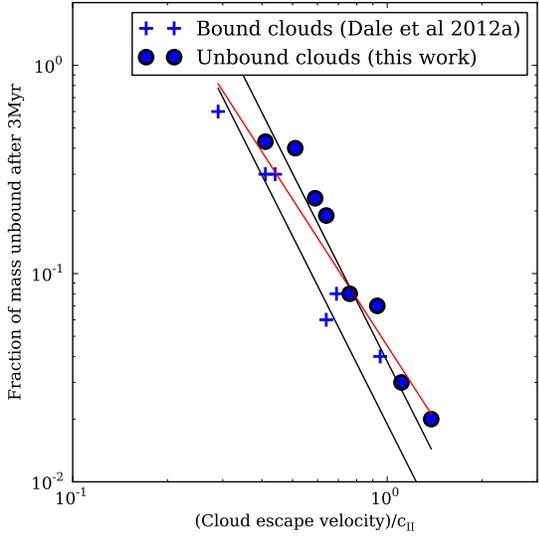}
\caption{Plot of unbound mass fractions against system escape velocities from the simulations presented in Paper I (blue crosses) and those presented here (blue circles). The red line is a fit to the simulation data. The black lines are the relation given in Equation \ref{eqn:fs2} computed assuming $f_{\rm i}=0.05$ (lower line) and $f_{\rm i}=0.10$ (upper line). Note that the escape velocities at the times when ionization was enabled have been used.}
\label{fig:fs_vesc}
\end{figure}

As in Paper I, we observe very strong differences across the parameter space in how much ionization influences the cloud dynamics. The more massive and denser clouds, e.g. Runs UC and UZ are largely unaffected, while the lower--mass and lower--density clouds such as Runs UQ and UF suffer substantial fractions of their gas reserves to be expelled. We note that the reaction of the clouds to photoionization again seems to depend strongly on their escape velocities. We therefore construct a simple hypothesis to explain this and compare it to the combined results of Paper I and this work.\\
\indent \cite{2012arXiv1208.3472B} recently presented an attractively simple model for determining whether a star--forming cloud would remain bound or not, in which the escape velocity of the cloud is compared with the sound speed in the ionized gas. This model works very well on the assumption that all the gas not involved in star formation is ionized on fairly short timescales, but this is not necessarily the case. In all simulations presented in Paper I and here, the fraction of gas which is ionized over the 3Myr timescale on which photoionization is permitted to act is low, typically a few to ten percent. However, in many simulations, the fractions of gas unbound from the model clouds is substantially larger than this. Clearly then, large quantities of gas are being unbound not because they are ionized themselves, but because they are being driven out of the cloud potential wells by the expanding HII regions. If the HII region is regarded as a piston pushing with a pressure $P_{\rm HII}$ against an area $\sim R_{\rm cloud}^{2}$, the impulse transmitted to the cloud over the timescale $t_{\rm SN}$ considered here will be $P_{\rm HII}R_{\rm cloud}^{2}t_{\rm SN}$. If the quantity of swept--up material has mass $M_{\rm s}=f_{\rm s}M_{\rm cloud}$ ($f_{\rm s}$ being the fractional swept--up mass) and velocity $v_{\rm s}$, we may write
\begin{eqnarray}
P_{\rm HII}R_{\rm cloud}^{2}t_{\rm SN}=f_{\rm s}M_{\rm cloud}v_{\rm s}.
\label{eqn:phii}
\end{eqnarray}
The pressure in the HII region may be estimated from the density and sound speed in the ionized gas:
\begin{eqnarray}
P_{\rm HII}=\rho_{\rm HII}c_{\rm HII}^{2}.
\label{eqn:phii2}
\end{eqnarray}
If we let the global ionization fraction be $f_{\rm i}$, then the total ionized mass $M_{\rm ionized}=f_{\rm i}M_{\rm cloud}$ and we may write approximately
\begin{eqnarray}
P_{\rm HII}\sim\frac{f_{\rm i}M_{\rm cloud}}{R_{\rm cloud}^{3}}c_{\rm HII}^{2}.
\end{eqnarray}
Inserting this into Equation \ref{eqn:phii}, we get
\begin{eqnarray}
f_{\rm i}\frac{t_{\rm SN}}{R_{\rm cloud}}c_{\rm HII}^{2}=f_{\rm s}v_{\rm s}.
\label{eqn:chii}
\end{eqnarray}
If we now take the limiting case where $v_{\rm s}=v_{\rm ESC}$, we may regard the swept--up mass as \emph{unbound mass} and rearrange for the unbound mass fraction $f_{\rm unbd}$:
\begin{eqnarray}
f_{\rm unbd}=f_{\rm i}\left(\frac{c_{\rm HII}}{v_{\rm ESC}}\right)\left(\frac{c_{\rm HII}t_{\rm SN}}{R_{\rm cloud}}\right).
\label{eqn:fs}
\end{eqnarray}
The two parenthetical terms on the right--hand side of Equation \ref{eqn:fs} have intuitive physical meanings. The first states that the unbound mass fraction is larger than the ionized gas fraction in proportion to the factor by which the ionized sound speed exceeds the system escape velocity, with higher escape velocities leading to smaller unbound mass fractions. The second term relates the size of the clouds to the distance that an HII--driven shell may travel within the imposed time interval before supernovae detonate, $t_{\rm SN}$, and states that, regardless of the escape velocity, larger clouds will be less affected by ionization simply because the shocks which it drives do not have time to explore the whole cloud volume.\\
\indent This phenomenon is particularly clear in the evolution of Run A from Paper I, in which ionization was able to blow a large number of well--cleared bubbles but, since their individual sizes are limited to $c_{\rm HII}t_{\rm SN}\approx 30$pc, was unable to disrupt a substantial fraction of that very large cloud. This is also clear to a lesser extent in Runs UB, UC and UU here where, although bubble formation is clearly in progress, large regions of the clouds are completely untouched.\\
\indent Equation \ref{eqn:fs} is very general but the clouds studied here and in Paper I obey the constraint obeyed by real star--forming clouds, as shown in the \cite{2009ApJ...699.1092H} sample, that their masses and radii are related by $M_{\rm cloud}\propto R_{\rm cloud}^{2}$, so that their column densities are all approximately the same. We may then write
\begin{eqnarray}
M_{\rm cloud}=AR_{\rm cloud}^{2},
\end{eqnarray}
where $A$ is a constant with the dimensions of a surface density whose value is approximately 0.03 g cm$^{-2}$. The cloud escape velocity $v_{\rm ESC}=\sqrt{(2GM_{\rm cloud}/R_{\rm cloud})}$ may then be written as
\begin{eqnarray}
v_{\rm ESC}=\sqrt{2GAR_{\rm cloud}}.
\end{eqnarray}
Rearranging and inserting this into Equation \ref{eqn:fs} yields
\begin{eqnarray}
f_{\rm unbd}=f_{\rm i}(2GAc_{\rm HII}^{2}t_{\rm SN})\left(\frac{1}{v_{\rm ESC}^{3}}\right).
\label{eqn:fs2}
\end{eqnarray}
We have already noted that the ionization fraction $f_{\rm i}$ in our simulations is small and does not vary very much, being 5--10$\%$ for almost all calculations presented here and in Paper I. There seems to be no relation between the values of $f_{\rm i}$ and the global cloud properties in these simulations. Treating $f_{\rm i}$ as a parameter therefore, we plot the relation given in Equation \ref{eqn:fs2} assuming values of $f_{\rm i}$ of 0.05 and 0.10, together with the measured values of $f_{\rm unbd}$ from Paper I and from the simulations presented here (where we again subtract the mass which becomes unbound in the corresponding control runs) in Figure \ref{fig:fs_vesc}.\\
\indent The red line in Figure \ref{fig:fs_vesc} is a linear fit to the logarithmically--plotted data and has a slope of -2.34, so that the simulation data imply that $f_{\rm unbd}\propto v_{\rm ESC}^{-2.34}$, which is somewhat flatter than the relation derived above in Equation \ref{eqn:fs2} of $f_{\rm unbd}\propto v_{\rm ESC}^{-3}$. However, the region between the fiducial black lines corresponding to $f_{\rm i}=0.05$ and $f_{\rm i}=0.10$ corresponds rather well with the results from the simulations. This implies that, even though it contains no information about the geometry of the clouds, Equation \ref{eqn:fs2} gives a reasonable fit to the simulation output within the range of observed ionization fractions.\\
\subsection{Unbinding embedded stars}
\begin{figure}
\includegraphics[width=0.45\textwidth]{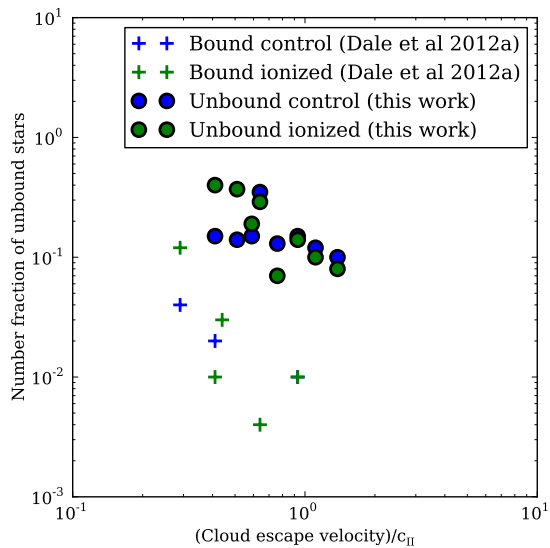}
\caption{Unbound fractions by number of stars/clusters in the unbound clouds (circles) and the bound clouds from Paper I (crosses). Blue and green symbols represent control and ionized simulations respectively.}
\label{fig:sunbndn}
\end{figure}

\begin{figure}
\includegraphics[width=0.45\textwidth]{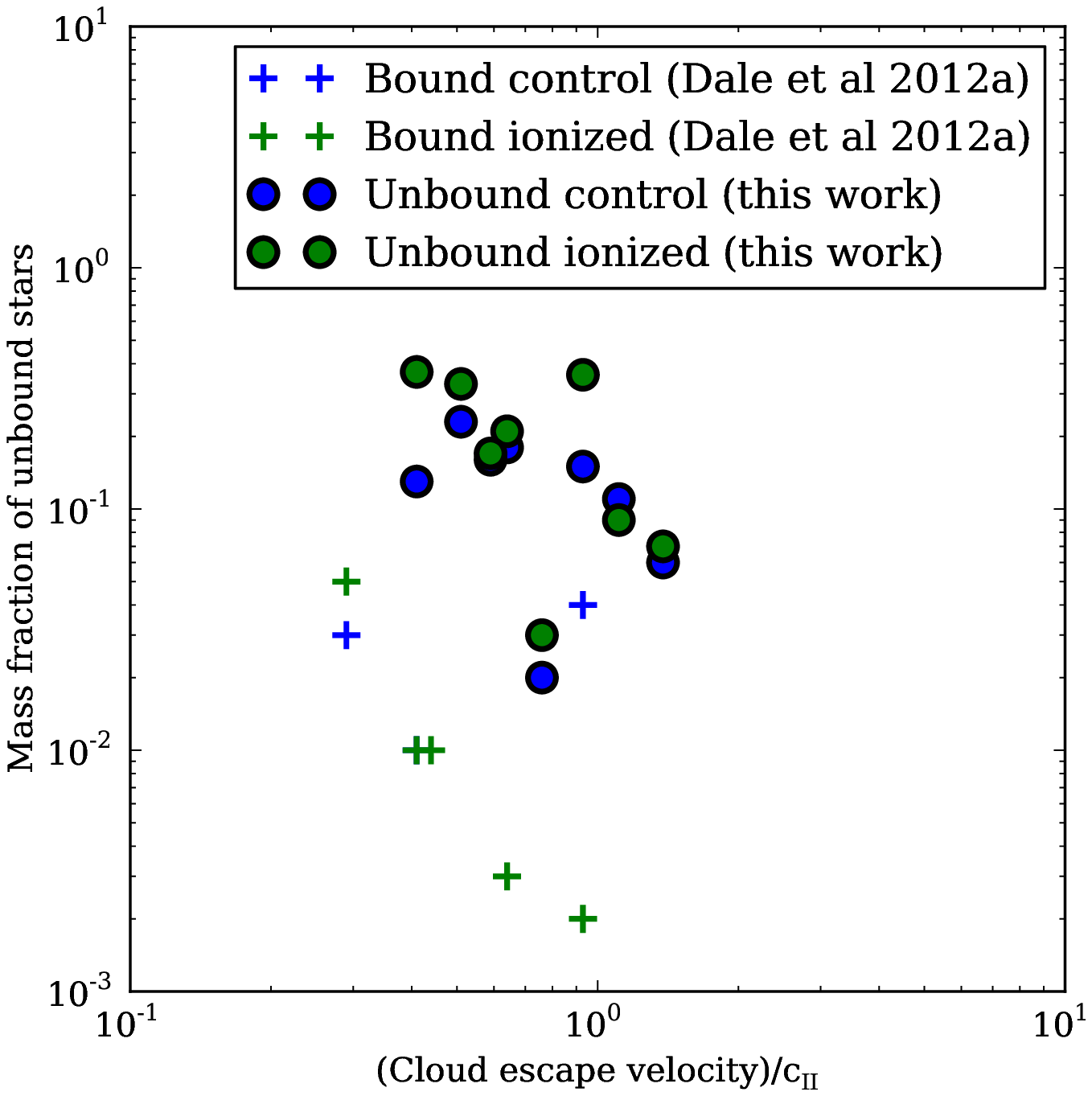}
\caption{Unbound fractions by mass of stars/clusters in the unbound clouds (circles) and the bound clouds from Paper I (crosses). Blue and green symbols represent control and ionized simulations respectively.}
\label{fig:sunbndm}
\end{figure}

\begin{figure}
\includegraphics[width=0.45\textwidth]{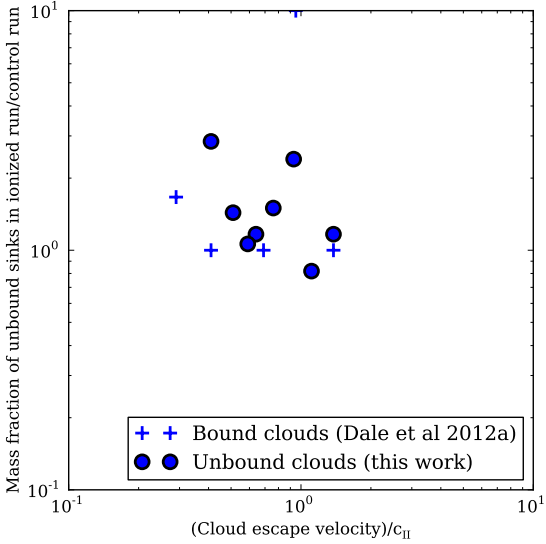}
\caption{Unbound fractions by number of stars/clusters in the unbound clouds (circles) and the bound clouds from Paper I (crosses). Blue and green symbols represent control and ionized simulations respectively.}
\label{fig:sunbndmcompare}
\end{figure}

\indent The discussion in the preceding section focusses on the expulsion of material from the potential wells of our model clouds, assuming all the unbound material to be in gaseous form. In reality, some of the unbound mass is actually unbound stars, although the contribution of stellar material to the unbound masses is small in almost all cases owing to the low fractions of gas converted to stars in most simulations. It is of great interest to measure the ability of feedback to expel stars from cloud potential wells. In Figures \ref{fig:sunbndn} and \ref{fig:sunbndm}, we show the fractions by number and mass respectively of stars unbound in all simulations from Paper I and presented here. Note that we again consider a sink to be unbound if it has positive total energy in the system centre of mass frame. We do not attempt to define bound subclusters which are nevertheless being ejected from the cloud. Blue symbols represent fractions of objects unbound in the control simulations, and green symbols fractions unbound in the ionized runs.\\
\indent We do not attempt to subtract off the unbound mass or number fractions from the control runs to isolate the effects of ionization here. Whether measuring mass or number fractions of stars unbound, there is considerably more scatter than when simply measuring total unbound mass fractions, as in the previous section. This is certainly due in part to small--number statistics since, in many cases, the numbers of unbound objects are small. In Figure \ref{fig:sunbndmcompare} we plot the mass fraction of unbound sinks in each ionized run divided by that in the corresponding control run. There is at best a weak correlation between this value and the escape velocity, implying that the dynamics of the stellar material is influenced by more than just the global system escape velocity.\\
\indent It is evidently not the case that the unbound stellar fraction can be simply related to the total unbound mass fraction by $f_{\rm unbd}^{*}=\epsilon f_{\rm unbnd}$, where $\epsilon$ is the star formation efficiency, and that the problem of estimating \emph{a priori} the unbound stellar mass fractions is much more difficult. In some simulation pairs, the unbound mass or number fraction in the control run is slightly \emph{greater} than in the ionized run, implying that simple modelling of this issue is very difficult due to the stochastic nature of the star formation process, particularly when stellar feedback is included. Nevertheless, the general conclusion may be drawn that photoionizing feedback does a relatively poor job of unbinding \emph{stellar mass} from clouds on the timescale studied here, even when it can be effective in unbinding gas. In no simulation presented here or in Paper I was photoionization able to unbind more than forty percent of the model clouds' stellar populations, whether measured by mass or number. This is despite total unbound mass fractions being much higher in some simulations (e.g. runs I and UQ). Most of the sinks which are unbound are not the result of the disruption of stellar systems, but are instead the result of star formation occurring in gas which was already unbound. We find that excluding the unbound gas from the calculation of the potential results in substantially more stars becoming unbound. The unbound fraction by number of stars in Run UQ rises to $\sim70$ percent if this is done, which indicates that although very large quantities of gas are unbound in this simulation (and others), this gas still dominates the gravitational potential, partly because there is so much of it, as the SFEs are generally low, and partly because feedback has not moved it very far in the $t_{\rm SN}$ time window.\\
\subsection{Escape of ionizing photons and supernova ejecta}
\begin{figure}
\includegraphics[width=0.45\textwidth]{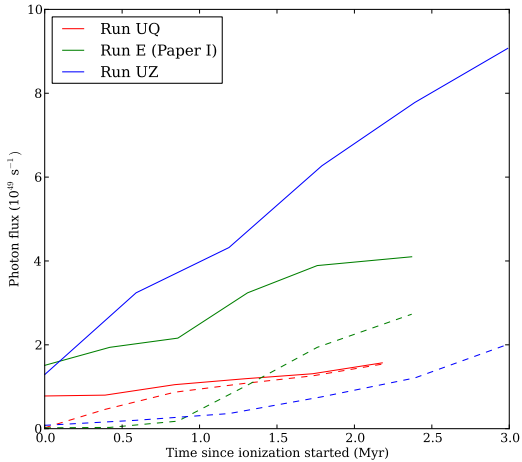}
\caption{Intrinsic (solid lines) and effective (dashed lines) ionizing photon fluxes for clouds UQ (red) and UZ (blue) from this work and for Run E (green) from Paper I, as functions of time since ionization started in each run.}
\label{fig:leak}
\end{figure}

\begin{figure}
\includegraphics[width=0.45\textwidth]{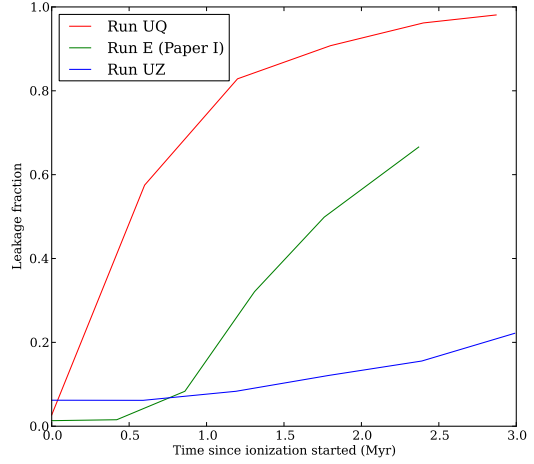}
\caption{Fractions of ionizing photons leaking from clouds UQ (red) and UZ (blue) from this work and for Run E (green) from Paper I, as functions of time since ionization started in each run.}
\label{fig:leak_666}
\end{figure}

As in Paper I, we analyse the clouds after the action of ionization to determine how permeable they are to ionizing photons and to the ejecta from the first supernova to detonate within. In Table \ref{tab:photons} we give the actual total luminosity of the stellar content of each simulation, the fraction of ionizing photons escaping the clouds, and their consequent effective ionizing luminosities. We reproduce the result from Paper I that clouds with higher intrinsic luminosities are proportionally less permeable to photons, so that the spread in effective ionizing luminosities is again small, almost all systems releasing $1$--$3\times10^{49}$ Lyman continuum photons per second into their surroundings. These numbers are of course computed at a single epoch -- the time at which the massive stars begin dying off. This is also, then, the time when the clouds' intrinsic ionizing luminosities might be expected to start decreasing, unless an increase in the star formation rate overproduces young massive stars and compensates for the older ones lost. In Figures \ref{fig:leak} and \ref{fig:leak_666}, we plot the intrinsic (solid lines) and effective (dashed lines) ionizing fluxes, and the escape fractions, all as functions of time, for three clouds (Runs UQ and UZ from this paper and Run E from Paper I). These three systems are representative of the behaviour seen in all simulated clouds across both papers.\\
\indent The intrinsic ionizing fluxes (Figure \ref{fig:leak}, solid lines) grow with time in all cases as star formation continues, the increase being fastest in Run UZ, where feedback has virtually no effect on star formation, slowest in Run UQ, where star formation is significantly affected, and intermediate in Run E. The leakage fractions (Figure \ref{fig:leak_666}) reflect the influence ionization has on the structure of each cloud. Run UQ is losing $80\%$ of its photons within $\sim1$Myr as the HII regions burst out of the cloud, whereas the leakage fraction in Run UZ grows very slowly as the ionized bubbles have much more difficulty breaking out of this cloud. Run E is again intermediate between these cases, its leakage fraction eventually growing to $\sim70\%$, but only after a delay of $\sim1$Myr during which it scarcely grows at all. These factors conspire together to make the effective ionizing fluxes (Figure \ref{fig:leak}, dashed lines) increase very slowly with time, so that for much of the $t_{\rm SN}$ time period studied here, the photon fluxes escaping all three clouds are the same to within a factor of a few.\\
\indent We have evolved our simulated clouds as close as possible to the time when the first supernovae are expected to detonate. The fate of the ejecta and the additional damage the first supernovae are likely to do to the clouds are given in Table \ref{tab:sn} in terms of the fraction of the ejecta from the first supernova which is likely to escape, the complementary fraction which is likely to be stopped by the cloud, and the quantity of mass the supernova is likely to unbind. The results of this analysis are similar to that presented in Paper I for the bound clouds, although there are some important differences. Firstly, all the unbound clouds will lose at least a few percent of the ejecta from the first supernova, which was not true of the most massive bound clouds. Secondly, one cloud (UQ) is likely to be be almost completely destroyed by the combined action of the photoionization and the first supernova explosion (at the end of the simulations presented here, the quantity of bound gas remaining in cloud UQ is $\approx1500$M$_{\odot}$, of which $\approx1200$M$_{\odot}$ will be unbound by the supernova, leaving $\approx300$M$_{\odot}$ still bound in the original centre--of--mass frame, about $3\%$ of the cloud mass).\\
\indent In addition, as we explain above, despite large quantities of gas being unbound in some of our simulations, this gas still exerts strong gravitational influence on the embedded clusters left behind, owing to its large mass and relative proximity to the sinks. Instantaneous removal of this gas to large radii would result in substantially larger fractions of sinks becoming unbound, which will likely occur in some simulations as a result of supernovae.\\ 
\begin{table}
\begin{tabular}{|l|l|l|l|}
Run & $Q_{\rm H}$ & f$_{\rm phot}$ & $Q_{\rm H}$' (10$^{49}$ s$^{-1}$)\\
\hline
UB&1.5&0.37&0.56\\
UC&4.5&0.50&2.3\\
UF&2.8&0.89&2.5\\
UP&2.9&0.98&2.8\\
UQ&1.6&0.98&1.5\\
UU&3.5&0.78&2.7\\
UV&1.3&0.85&1.1\\
UZ&9.1&0.23&2.1\\
\end{tabular}
\caption{Total ionizing luminosities, estimated fraction of ionizing photons leaking from clouds, and their consequent estimated effective ionizing photon luminosities.}
\label{tab:photons}
\end{table}
\begin{table}
\begin{tabular}{|l|l|l|l|}
Run & f$_{\rm esc}$ & f$_{\rm stop}$ & M$_{\rm unbnd}$ (M$_{\odot}$)\\
\hline
UB&0.02&0.98&43\\
UC&0.13&0.87&125\\
UF&0.56&0.44&1272\\
UP&0.76&0.24&961\\
UQ&0.77&0.23&1207\\
UU&0.13&0.87&229\\
UV&0.05&0.95&120\\
UZ&0.04&0.94&31\\
\end{tabular}
\caption{Estimated fractions of the the ejecta of single supernovae escaping from and being retained by clouds, and of the mass unbound by the first supernova, after the action of photoionization.}
\label{tab:sn}
\end{table}
\section{Discussion}
\indent The results of this study are complementary and comparable to those of Paper I. We find that the ability of photoionization to expel gas from star--forming clouds is a very strong function of the escape velocities of the clouds, regardless of whether the clouds are initially bound with virial ratios of 0.7 or, as here, formally unbound with virial ratios of 2.3. Subtracting off the fraction of gas which is unbound whether feedback is acting or not, we find that ionization is able to expel several tens of percent from lower--mass (1-3$\times10^{4}$M$_{\odot}$ clouds and a few tens of percent of the gas mass from larger, and therefore low escape velocity 10$^{5}$M$_{\odot}$ clouds. However, it is largely unable to expel material from denser 10$^{5}$M$_{\odot}$ clouds and those with larger masses. The fact that the clouds were partially unbound initially does not aid ionization in destroying them because the ionizing sources are, of course, located in the denser parts of the clouds which are bound.\\
\indent Drawing together results briefly alluded to in Paper I and combining with those discussed here, we find that photoionization is not very good at unbinding the stellar component of embedded clusters, particularly when the progenitor clouds are initially bound. Despite considerable scatter in both the mass and number fractions of unbound objects (Figures \ref{fig:sunbndn} and \ref{fig:sunbndm}), it is clear that the increase in the fraction of stars/subclusters unbound in the simulations presented here is only modestly increased (and occasionally decreased in fact) by the action of feedback. This occurs for reasons similar to those stated in the preceding paragraph; stars form in the densest regions of gas and are often well on the way to forming bound, virialised and even gas--free systems before any of the stars/subclusters become massive enough to exert feedback.\\ 
\indent Comparing Figure \ref{fig:gallery_final} here to Figure 8 from Paper I, the main difference in gas morphology between the unbound and bound clouds is the greater prevalence of a bubble--like morphology in the simulations presented here. All but run UZ show various numbers of usually interconnected bubbles cleared to a greater or lesser degree of gas. In contrast, Run A in Paper I shows a small number of largely separated bubbles, and only runs D, I and J show the kind of well--cleared bubbles prevalent in this work. Starting from partially unbound clouds does, therefore, assist photoionization in blowing bubbles.\\
\indent Comparison of the morphology of these simulations with that of real objects is irresistible, particularly in view of the superb infrared data being produced by telescopes such as Spitzer \citep[e.g.][]{2006ApJ...649..759C,2007ApJ...670..428C} and WISE \citep{2012ApJ...744..130K}. While these simulations imply that it is not as efficient at unbinding star forming regions as perhaps thought, photoionization is good at creating large evacuated bubbles around clusters. The corollary of these statements is of course that the presence of large evacuated bubbles around clusters should not be taken to mean that the clusters have been unbound, or even strongly affected, by feedback. However, detailed quantitative comparison is difficult for two main reasons.\\
\indent The majority of observed bubbles lie at unknown distances, so their basic physical parameters cannot be derived. While \cite{2006ApJ...649..759C} and \cite{2007ApJ...670..428C} catalogue almost 600 infrared bubbles in total, they are able to give distances (and therefore true sizes) for less than a hundred, and most of those suffer from the near/far distance degeneracy. If their far distances are accepted, about two thirds, and if the near distances are accepted, almost all, bubbles in these papers have radii less than 10pc. This would seem to present a problem, since most of the bubbles produced in our calculations have radii of this order for most of the time (although they are obviously smaller when they are younger). However, \cite{2006ApJ...649..759C} point out that larger bubbles are fainter and therefore more difficult to detect. The bubbles reported by \cite{2012ApJ...744..130K} are larger, with most having radii in the range 10--20pc.\\
\indent However, even with the best available data, the simulations are not yet in a state where they can be quantitatively compared to real bubbles. The images presented here are simple--minded column--density maps of all gas and are surely not what the model clouds would look like to an observer if they were real. It is possible to post--process simulation output and produce maps of ionized emission lines \citep[e.g][]{2012MNRAS.420..141E}, dust emission \citep{2011arXiv1109.3478W} or molecular lines (if abundances of the relevant molecules can be assumed, \cite{2011ApJ...743...91O}). However, it is clearly desirable to model the radiation transfer and chemistry correctly during the dynamical simulations themselves \citep[e.g][]{2012arXiv1205.6993H,2012MNRAS.421....9G}, despite the computational expense.\\
\indent There are several important caveats to this work, such as neglect of jets and magnetic fields, which are discussed in some detail in the latter sections of Paper I. However, the most important may be the fact that we do not include winds, and we here expand on this topic.\\
\indent As well as strong ionizing fluxes, O--stars also usually possess powerful winds. However, the combined effects of both these forms of feedback on an inhomogeneous circumstellar medium is not well understood. \cite{2001PASP..113..677C} suggest that, for reasonable ionizing fluxes, wind mass loss rates and terminal velocities, the circumstellar gas must be extremely dense in order that the effect of the HII region is overwhelmed by that of the wind. However, the physics of stellar wind interaction with either a previously--existing HII bubble or with the bare cool ISM are exceedingly complex and poorly understood, depending as they do on microphysical processes such as thermal conduction and mixing. Two--dimensional simulations by \cite{2003ApJ...594..888F} and \cite{2006ApJ...638..262F} of the combined impact of the HII region and wind from (respectively) a 35 and a 60M$_{\odot}$ show complex morphology despite the uniformity of the ambient medium used. In the case of the 65M$_{\odot}$ object, the wind rapidly sweeps the HII region into a thin shell, although the total kinetic energy imparted by the combined feedback mechanisms is only a factor of four larger than that provided by the HII region acting alone. In the case of the lower mass star with its weaker wind, the reverse shock between the wind bubble and HII region ceases to be supersonic after only a few$\times10^{5}$ yr and the HII region is swept into a much broader shell and the energetic effect of the wind is necessarily smaller. The authors caution that they have neglected thermal conduction, which may cause the wind bubbles to cool earlier, but also magnetic fields, which may mitigate this effect somewhat. The combined effects of HII regions and winds on an initially inhomogeneous medium are still very unclear.\\
\indent From an observational point of view, the picture is scarcely clearer. If winds are important in the evolution of very young star--forming regions, such systems should be full of diffuse gas with temperatures of around 10MK which should be visible in X--rays. Such hot diffuse gas is indeed observed in some star--forming regions e.g. M17, NGC 3603, NGC 3576 \citep{2003ApJ...593..874T,2006astro.ph..8173T}. If winds are important in this context, the photoionized HII region should be compressed by the wind bubble into a thin shell lining the feedback blown cavity, as seems to be the case in M17, and there should also be evidence of very hot wind gas interacting directly with cold molecular material \citep{2011ApJS..194...16T}. However, diffuse X--ray emission is absent from other systems, such as Orion, and the Eagle and Lagoon nebulae \citep{2003ApJ...593..874T}, and observations by \cite{2008IAUS..246...55S} of NGC 346 in the Small Magellanic Cloud reveal no evidence of any strong motions within the ionized gas and the dynamics of this region are attributed purely to classical HII region expansion. This may be a consequence of the low metallicity and hence more feeble stellar winds in the SMC. In any case, this suggests that there are at least some regions where the feedback--driven evolution is almost entirely due to photoionization.\\
\section{Conclusions}
\indent We have presented the results of a suite of simulations intended to see whether photoionizing feedback from O--type stars acting alone for the 3Myr timescale before the stars are expected to supernova is able to efficiently disperse clouds and embedded clusters. Synthesising the results of this work and Paper I, we have reached the following general conclusions:\\
\indent (1) The ability of ionization to disrupt clouds is a very strong function of the cloud escape velocities, owing to three facts, namely that the sound speed in the ionized gas is fixed at $c_{\rm HII}\approx10$km s$^{-1}$, only a small fraction of gas is likely to be ionized, and that the HII region therefore acts like a piston trying to push neutral gas out of the clouds' potential wells.\\
\indent (2) Massive clouds ($M_{\rm cloud}\gtrsim10^{5}$M$_{\odot}$) with realistic radii mostly have escape velocities comparable to or greater than $c_{\rm HII}$ and are thus dynamically immune to the effects of photoionization, regardless of whether the clouds are initially bound.\\
\indent (3) Lower mass clouds ($M_{\rm cloud}\sim10^{4}$-$10^{5}$M$_{\odot}$) with realistic radii have escape velocities considerably less than $c_{\rm HII}$ and are likely to lose several tens of percent of their gas in the 3Myr time window, regardless of whether the clouds are initially bound.\\
\indent (4) Photoionization alone is not an efficient means of unbinding the stellar component of embedded clusters, particularly if the progenitor clouds are bound initially. Even in low--mass and low--escape velocity clouds, ionization struggles to unbind a few tens of percent (by number or mass) of the stellar population, even when significant fractions of the cloud and stars are initially unbound.\\
\indent (5) The formation of pronounced and well--cleared bubbles on 3Myr timescales is relatively rare in bound clouds, occurring only in low--mass and/or very low--density clouds. Conversely, bubble formation is almost ubiquitous in unbound (and therefore intrinsically expanding) clouds, occurring in all but the most massive and most strongly bound objects. However, the presence of well--cleared bubbles does not imply that the host cloud has been strongly dynamically influenced by feedback.\\
\indent (6) Star formation efficiencies in the unbound clouds are generally lower than in the bound clouds regardless of the action of ionizing feedback. Photoionization in all cases reduces the star formation efficiency over the timescales studied here, but never by very large factors, particularly in the cases where the clouds are already partially unbound. In the unbound clouds, star formation efficiencies with or without feedback are a few to ten percent in most cases, reaching 20--25$\%$ in denser clouds.\\
\indent (7) The more massive clouds have, as expected, a more massive stellar population and therefore higher total ionizing fluxes, but they are proportionately less permeable to ionizing photons. Total photons luminosities range from a little more than $1\times10^{49}$ s$^{-1}$ for the low--mass clouds to nearly $10^{50}$ s$^{-1}$ for the most massive and most populous cloud, but photon escape fractions vary inversely with cloud mass from a few tens of percent to almost 100$\%$. We find that almost all systems studied here have effective ionizing luminosities into the intercloud medium in the range 1-3$\times10^{49}$s$^{-1}$, and that these luminosities vary rather slowly in time.\\
\indent (8) Almost all the clouds studied, whether bound or unbound are likely to survive their first supernova explosion and are therefore likely to trap some fraction of its ejecta to be involved potentially in another round of metal--enriched star formation. Conversely, the unbound clouds and the lower--mass bound clouds are all leaky to supernova ejecta to some degree and will lose form a few percent to more than three quarters of the ejecta from their first supernova directly to the intercloud medium.
\section{Acknowledgments}
We thank the referee, Mordecai--Mark Mac Low, for a careful reading of the manuscript and insightful suggestions regarding turbulence. This research was supported by the DFG cluster of excellence ?Origin and Structure of the Universe? (JED, BE).

\bibliography{myrefs}

\label{lastpage}

\end{document}